\documentclass[psamsfonts]{amsart}
\usepackage[utf8]{inputenc}
\usepackage{amsmath}
\usepackage{amsfonts}
\usepackage{amssymb}
\usepackage{enumerate}
\usepackage{mathrsfs}

\title{Exploring Weak Strategy-Proofness in Voting Theory}
\author{Anne Elizabeth Carlstein}
\date{March 2020}

\begin{document}

\begin{titlepage}
    \begin{center}
        \vspace*{1cm}
            
        \Huge
        \textbf{Exploring \\ Weak Strategy-Proofness \\ in Voting Theory}
            
        \vspace{0.8cm}
        \LARGE
        Anne Elizabeth Carlstein
            
        \vspace{0.8cm}
            
        \textbf{Advisor:  Eric Maskin}
            
        \vspace{8cm}
            
        A thesis submitted in partial fulfillment of the requirements for the degree of Bachelor of Arts with Honors
in the Mathematics Department\\
            
        \vspace{0.8cm}

        Harvard University
        
        \vspace{0.8cm}
        
        \Large

        March 29, 2020
            
    \end{center}
\end{titlepage}

\tableofcontents

\newpage
\section*{Introduction}

In social choice theory, the intent is to choose a winning alternative, given preferences of voters.  For the purposes of this thesis, we are only considering a choice between three alternatives $\{x, y, z\}$.  To understand how this is accomplished, we must first determine how voters and their preferences are expressed.  Here, we define a voter (denoted $i$) as a member of a continuum\footnote{This will ensure that the chances of a tie are negligible, i.e., we say that ties are nongeneric.}, drawn from the unit interval (i.e., $i \in [0, 1]$).  This lets us measure the proportion of voters with various preference rankings.  A voter's preferences are expressed as an ordinal ranking of the alternatives (where $x \succ y$ means that option $x$ is preferred to option $y$).

A domain (denoted by $U$) is a collection of preference rankings, each of which we express as a column vector.  For example, consider the following domain:

$$U =  \Bigg\{ \begin{tabular}{ c c c }
 x  & y & y \\ 
y  & x & z  \\  
z  & z & x   
\end{tabular} \Bigg\} $$
The third column vector $\begin{pmatrix} y \\ z\\ x \end{pmatrix}$ represents the preference ranking $y \succ z \succ x$.
With a specific profile (denoted by $u$), we can represent the relative proportion of voters and their preferences about the alternatives.  For example, consider the following profile:

$$ u = \begin{tabular}{ c c c }
p & q & 1 - p - q \\
\hline
 x  & y & y \\ 
y  & x & z \\  
z  & z & x     
\end{tabular}$$
In this profile, a proportion $q$ of the voters have preference ranking $\begin{pmatrix} y \\ x \\ z \end{pmatrix}$.  We can similarly identify the other preference rankings and their weights in the profile.

Given a preference ranking of this form, how would we determine the winning outcome?  This can be accomplished by a voting rule.  For the purposes of this paper, given a profile of voter preferences $u$, and a collection of alternatives $X$, a \textbf{voting rule} (denoted $F$) selects a winner (if one exists).  More formally, given a set of alternatives $X \subseteq \{x, y, z\}$, a voting rule $F$ maps each profile $u$ on domain $U$ to an alternative in $X$.  This is denoted $F(u, X)$.\footnote{This definition only requires the voting rule to map to a winning alternative on generic profiles.  On a nongeneric profile, a voting rule can evaluate to $\varnothing$.}

Given this formulation of a voting rule, we still have not specified how the voting rule actually selects the result.  There are numerous methods to choose a winner.  We would not be happy with a voting rule that says ``given options $\{x, y, z\}$, $x$ is always the winner, regardless of voter preferences."  (For example, a Russian election where Vladimir Putin is on the ballot).  So, we want voting rules to satisfy certain desirable axioms.  For example, consider the profile $u$ that we defined above.  Intuitively, if a voting rule selected $z$ as the winner, we might be dissatisfied with that result, because clearly, everyone prefers option $y$ over option $z$.  (This is the Pareto property, as is defined below).

Certain axioms are extremely natural:

\textbf{Pareto} (denoted P) - As we saw before, if one alternative ($y$) is preferred by everyone to another alternative ($z$), then $z$ should not be the alternative that is selected by the voting rule.  It also follows from this axiom that if everyone ranks alternative $y$ above every other option, then $y$ should be the winner.

\textbf{Anonymity} (denoted A) - This tells us that if the names of voters are relabeled, then the outcome of the voting rule should not change.  To express this more formally, we take some measure preserving permutation $\pi: [0, 1] \to [0, 1]$ (i.e., for any $C \subseteq [0, 1], \mu(C) = \mu(\pi(C))$).  Apply permutation $\pi$ to all voters $i$ in some profile $u$, so that each voter $i$ has voter $\pi(i)$'s preference rankings, and denote the fully permuted profile as $u_\pi$.  To satisfy this property, it must be true that $F(u, X) = F(u_\pi, X)$.

\textbf{Neutrality} (denoted N) - If we permute the names of the candidates on the ballot (and accordingly adjust the voter preferences associated to them), then the adjusted winner should be the permuted winning candidate.

For example, consider the following permutation on profile $u$:

$$\sigma(x) = y; \sigma(y) = z; \sigma(z) = x$$

$$ u_{\sigma} = \begin{tabular}{ c c c }
p & q & 1 - p - q \\
\hline
 y  & z & z \\ 
z  & y & x \\  
x  & x & y    
\end{tabular}$$

Suppose that the winner according to some voting rule $F(u, X) = y$.  So then, to satisfy Neutrality, it should be true that $F(u_{\sigma}, X) = \sigma(y) = z$.

The above axioms are clearly ones that would be attractive for a voting rule to fulfill.  However, some other axioms in the voting literature are less obvious.  In particular, the next axiom is particularly controversial:

\textbf{Independence of Irrelevant Alternatives} (denoted IIA) - Let $X$ be a collection of alternatives, and let $X'$ be a collection of alternatives where $x \in X' \subset X$.  If alternative $x$ is the winner $F(u_., X)$, then to satisfy IIA, $x$ must also be the winner $F(u_., X')$.  In other words, dropping non-winning candidates from consideration should not change who the winner is.  

Surprisingly, these axioms are not easy to fulfill.  This is apparent in the most celebrated result of social choice theory, Arrow's Impossibility Theorem (Arrow 1951).

\textit{\textbf{Theorem (Arrow's Impossibility Theorem, as stated in Dasgupta and Maskin 2019)} Assume the domain $U$ consists of all possible preference rankings over $\{x, y, z\}$. There is no voting rule that satisfies P, A, N, and IIA.}\footnote{This theorem applies to all sets of $3$ or more alternatives, but for the purposes of this thesis, we are focusing on just the case of $3$ alternatives.}

This result might seem disheartening - however, beyond this result, there is still a wide variety of directions that voting theory can take.  We can consider how to restrict the domain of preferences.  We can reconsider which axioms are included.  With these modifications, we can analyze how these restrictions limit the voting rules that are possible.  And that is what we will do!

First, we introduce some of the common voting rules:

\textbf{Rank-Order Voting a.k.a. Borda Count (Borda 1781)} - The alternatives are ranked by each voter (in the order of their preferences), and the alternatives are scored according to the following rule:  an alternative gets $k$ points for each voter that ranks it above $k$ other alternatives.  The highest scoring alternative wins.  This voting method dates back to 1781, and was proposed by Jean-Charles Borda.  

\textbf{Majority Rule a.k.a. Condorcet's Method (Condorcet 1785)} - Again, the alternatives are ranked by each voter.  Each pair of alternatives are compared in a head-to-head contest (aggregating over all voters), and the winning option is the alternative $x$ where a majority of voters prefer $x$ over any other opponent $y$.  This method also dates back to the 18$^{th}$ century (1785), and was championed by the Marquis de Condorcet.  (Borda and Condorcet were contemporaries and intellectual foes - they disagreed on which voting rule was superior.)

\textbf{Plurality Rule} - Given the preference rankings of each voter, the alternative that is ranked in the first spot by the most voters is declared the winner (even if they are short of a majority)\footnote{This definition of plurality rule makes sense given the preference rankings of voters - however, in practice, plurality rule does not actually require voters to rank all alternatives (they only need to report their one top choice).}.  This method is used in U.S. elections (e.g., to elect members of Congress, to determine the winner of each state's electoral votes, for many local elections).

Of course, there are many voting rules besides the methods described above.  However, in the spirit of Arrow's Impossibility Theorem, Dasgupta and Maskin (2019) show in their \textbf{Theorem 6} that the Borda Count and Condorcet's Method are the only possible voting rules when certain intuitive axioms are imposed.  This thesis generalizes Dasgupta and Maskin's \textbf{Theorem 6} by relaxing a key axiom.  

In \textbf{Theorem 6}, attention is restricted to \textbf{rich domains:}

A domain $U$ is a \textbf{rich domain} if for each $x \in X$ there exists some $y, z \in X$ and some preference ranking in the domain such that $y \succ x \succ z$.  In other words, each alternative must appear in a non-extremal place in some preference ranking in the domain.

For example, consider the following domains:

$$ U^1 = \Bigg\{ \begin{tabular}{ c c c c}
 x  & x & y & z\\ 
y  & z & z & y \\  
z  & y & x &x    
\end{tabular} \Bigg\} \quad  U^2 = \Bigg\{ \begin{tabular}{ c c c c}
 x  & x & y & y\\ 
y  & z & z & x\\  
z  & y & x & z    
\end{tabular} \Bigg\}$$
(Clearly, $U^1$ is not rich, while $U^2$ is). 

Additionally, their \textbf{Theorem 6} replaces the common IIA axiom with \textbf{Strategy-Proofness}.

\textbf{Strategy-Proofness} (denoted SP) - 
Consider a generic profile $u$ on the domain $U$, and alternatives $X$, and suppose that $F(u, X) = x$.  Let the notation $x \succ _i y$ denote that voter $i$ prefers alternative $x$ over alternative $y$.  Consider all coalitions $C \subseteq [0, 1]$, and all profiles $u'$ where for all $i \notin C$, $\succ_i$ $=$ $\succ_i '$. Then, for $y = F(u', X)$, there exists some $i \in C$ such that $ x \succ _i y $. 

Suppose that $u$ represents the true preferences of the voters.  For any strategic profile $u'$, where members of the coalition $C$ misreport their preferences, there is some voter in the misreporting coalition who winds up with a worse result than what they would have received had they reported their true preferences.  Or, in other words, if a coalition of voters can misreport and improve their outcome, then the voting rule is not Strategy Proof.  Clearly, this is an appealing axiom, as we want voters to be incentivized to report their true preference rankings, and to avoid attempting to (potentially misguidedly) game the system for a better result.

Before stating Dasgupta and Maskin's \textbf{Theorem 6}, we formally define the two voting rules, Condorcet's Method (denoted $F^C$), and Borda Count (denoted $F^B$):

\textbf{Majority Rule (a.k.a. Condorcet's Method)} - 

$$ F^C(u, X) = \{ x \in X | \mu[i | x \succ _i y ] \geq \frac{1}{2} \forall y \neq x, y \in X \}$$

In other words, when considering the head-to-head comparisons of all pairs, the Condorcet winner $x$ is the one for which a majority prefer $x$ to every other alternative $y$. 

Recall profile $u$ that we defined above:

$$ u = \begin{tabular}{ c c c }
p & q & 1 - p - q \\
\hline
 x  & y & y \\ 
y  & x & z \\  
z  & z & x     
\end{tabular}$$

For example, apply $F^C$ to profile $u$.  We see that in the pairing of $(x, y)$, $x$ is preferred to $y$ with proportion $p$.  And, in the comparison $(x, z)$, $x$ is preferred to $z$ with proportion $p + q$.  Finally, in the comparison $(y, z)$, $y$ is always preferred.  This tells us that $z$ can never be the Condorcet winner.  We also can see that if $p > \frac{1}{2}$, then $x$ is the Condorcet winner.

However, note that Condorcet's Method does not always select a winner.  Consider the following profile:

$$ u_c = \begin{tabular}{ c c c }
1/3 & 1/3 & 1/3 \\
\hline
 x  & y & z \\ 
y  & z & x \\  
z  & x & y     
\end{tabular}$$

We apply $F^C$ to profile $u_c$.  We can see that a proportion $\frac{2}{3}$ prefer $x$ to $y$; a proportion $\frac{2}{3}$ prefer $y$ to $z$, and a proportion $\frac{2}{3}$ prefer $z$ to $x$.  This is a Condorcet Cycle.  According the head-to-head contests, we have that the overall electorate prefers $x \succ y$, $y \succ z$, and $z \succ x$.  This means that no one can be the Condorcet winner - giving us the Condorcet paradox (as noted by the Marquis de Condorcet).

\textbf{Ranked-Choice Voting (a.k.a. Borda Count) -}

$$ F^B(u, X) = \{ x \in X | \int r_i(x) d\mu(i) \geq \int r_i(y) d\mu(i) \quad \forall y \in X \} $$
(Here, the point score $r_i(x) = |\{ y \in X | x \succeq _i y \}| $)

The score for each alternative is determined by integrating over the point scores $r_i$ for all voters $i$.  The point score $r_i(x)$ is measured as the number of candidates that are ranked no higher than $x$ by voter $i$.  The Borda winner is the alternative whose total score is higher than the score of any other candidate.

As an example, we can apply the Borda Count to profile $u$:

The score for alternative $x$ is:  $2p + q$

The score for alternative $y$ is:  $p + 2q + 2(1 - p - q) = 2 - p$.

The score for alternative $z$ is: $1 - p - q$.

Note that $z$ can never be the winner under the Borda Count because this would require that $1 - p - q > 2 - p$, which would imply that $q < -1$, which is impossible.
Also note that $p > \frac{1}{2}$ is not enough to guarantee that $x$ is the Borda winner, even though $p > \frac{1}{2}$ guaranteed that $x$ would be the Condorcet winner.

So, with this setup, we can state the theorem:

\textit{\textbf{Theorem 6 (Dasgupta and Maskin)}  If $F$ satisfies P, A, N, and SP on U, and U is rich, then $F = F^B$ or $F = F^C$.}

The proof of this theorem proceeds in three steps.  In the first step of the argument, it is shown that the Borda Count is the only voting rule that satisfies the axioms on the \textbf{Condorcet Cycle} domain.  

A Condorcet Cycle domain (on 3 alternatives) is:

$$ U^{CC} = \Bigg\{ \begin{tabular}{ c c c c}
 x  & y & z \\ 
y  & z & x \\  
z  & x & y    
\end{tabular} \Bigg\}$$

Next, for the second step, it is proven that no voting rule satisfies the axioms on any expansion of the Condorcet Cycle domain.  The final step shows that Condorcet's Method is the only voting rule that satisfies the axioms on any remaining rich domain.

Richness is an important component of this result.  Without this domain restriction, there are non-rich domains where a voting rule other than $F^C$ and $F^B$ satisfy the given axioms.   Additionally, in the proof, richness also allows for permuting the alternatives (applying axioms A and N), while still remaining in the same domain.

Returning to the definition of Strategy-Proofness, we notice that there is no limit on the size of the coalition that deviates.  However, manipulation on a large scale is unrealistic in practice, so instead, we focus only on manipulations by coalitions of a small size ($\varepsilon$).  This allows us to formulate a less restrictive axiom, namely  \textbf{Weak Strategy-Proofness} (as suggested by Shengwu Li, and as defined by Dasgupta and Maskin 2019).

A voting rule is \textbf{manipulable} on $U$ if for all $\varepsilon > 0$, there exist some coalition $C$ where $|C| < \varepsilon$, profiles $u$ and $u'$ (where for all $i \notin C$, $\succ_i$ $=$ $\succ_i '$) and $x, y \in X$, where $x = F(u, X), y = F(u', X)$, and $y \succ_i x$ for all $i \in C$.

\textbf{Weak Strategy-Proofness} (denoted WSP) - A voting rule $F$ satisfies Weak Strategy-Proofness on $U$ if $F$ is not manipulable on $U$.

In other words, if a coalition of arbitrarily small size can misreport their preferences and achieve an improved result, then the voting rule is manipulable.

Because Weak Strategy-Proofness allows for a larger class of voting rules to be considered, when taken as an axiom, this leads to a generalization of Dasgupta and Maskin's \textbf{Theorem 6}.

So, for my senior thesis, I prove the following result:

\textit{\textbf{Theorem}  If $F$ satisfies P, A, N, and WSP on U, and U is rich, then $F = F^B$ or $F = F^C$.}

This proof proceeds in three steps, as in the proof of \textbf{Theorem 6}.  However, this proof requires many more subcases, to check for the possibility of small group misrepresentations.  This utilizes a partitioning argument (based on $\varepsilon$), which is a generalization of the argument used in the proof of \textbf{Theorem 5} (Dasgupta and Maskin).  The proof also introduces iterative and inductive machinery to relate the evaluation of profiles that differ by more than $\varepsilon$.  

Without further ado, we begin the proof.

\newpage

\section*{Step 1}
First, we want to show that the Borda Count is the unique voting rule that satisfies P, A, N, and WSP on the Condorcet Domain:

$$ U^{CC} = \Bigg\{ \begin{tabular}{ c c c c}
 x  & y & z \\ 
y  & z & x \\  
z  & x & y     
\end{tabular} \Bigg\} $$

Because WSP is less restrictive than SP, we know from Barbie et al. 2006 that $F^B$ also satisfies WSP on the Condorcet Domain.  (By properties of the Borda Count, $F^{B}$ always satisfies P, A, N).

We consider the following profile:

$$ u_. = \begin{tabular}{ c c c }
a &  b  & 1 -  a - b\\
\hline
 x  & y  & z\\ 
y  & z & x\\  
z  & x &y   
\end{tabular}$$

Since $U^{CC}$ is symmetric, for this entire step, we can assume without loss of generality that $x$ is the winner according to the Borda Count.  
This means that:

$$2a + (1 - a - b) > a + 2b $$

$$ 2a + (1 - a - b) > b + 2(1 - a -b)$$

Assume that $F$ is some voting rule that satisfies P, A, N, and WSP, and that $F \neq F^B$.  Fix $\varepsilon > 0$.

\textbf{Step 1, Case I:} Suppose that according to the Borda Count, $y$ beats $z$. 

So, we have the following inequalities:

$$ a + 1 - b > a + 2b > 2 - 2a -b$$

$$ a + b > \frac{2}{3} $$

$$ a > \frac{1}{3} > b$$

\textbf{1.I.1} Assume that $F$ has the following evaluation:

$$F(u.) = y \quad (A1)$$

Consider the permutation $\sigma$ on $u.$, and the resulting profile, where $\sigma(x) = z, \sigma(y) = x, \sigma(z) = y$:

$$ u_1 =\begin{tabular}{ c c c }
a &  b  & 1 -  a - b\\
\hline
 z  & x  & y\\ 
x  & y & z\\  
y  & z &x   
\end{tabular} = \begin{tabular}{ c c c }
b &  1 - a - b  & a\\
\hline
 x  & y  & z\\ 
y  & z & x\\  
z  & x &y   
\end{tabular} $$

By A and N, we have:

$$ F(u_1) = x \quad (A2)$$

\textbf{1.I.1.1}:  Suppose that $1 - a - b \geq b$

\textbf{Case 1.I.1.1.1} $0 \leq a - b < \varepsilon$

In this case, voters in profile $u_.$ with preferences $\begin{pmatrix} x \\ y \\ z \end{pmatrix}$ will want to improve their outcome.  A weight of $a - (1 - a - b)$ will misreport as $\begin{pmatrix} z \\ x \\ y \end{pmatrix}$, and a weight of $(1 - a - b) - b$ will misreport as $\begin{pmatrix} y \\ z \\ x \end{pmatrix}$.  This will have a total coalition size of $2a - 1 + b + 1 - a - 2b = a - b < \varepsilon$.  It will induce profile $u_1$ and evaluation $(A2)$, giving improved result $x$.  This would contradict WSP, so \textbf{Case 1.I.1.1.1} can not hold.  

\textbf{Case 1.I.1.1.2} $\varepsilon \leq a - b < 2\varepsilon$

Consider the following profile:

$$ u_2 =\begin{tabular}{ c c c }
b + (k + m) &   1 - a - b - k  & a - m\\
\hline
 x  & y  & z\\ 
y  & z & x\\  
z  & x & y  
\end{tabular} $$

Where:  $k = \frac{1 - a - 2b}{2}$ ; $m = \frac{2a + b - 1}{2}$

Suppose that $F(u_2) \neq x$.  But then, note that voters in profile $u_2$ with preferences $\begin{pmatrix} x \\ y \\ z \end{pmatrix}$ would form a coalition of size $k + m$ and misreport as $\begin{pmatrix} y \\ z \\ x \end{pmatrix}$ with a weight of $k$, and would misreport as $\begin{pmatrix} z\\ x \\ y \end{pmatrix}$ with a weight of $m$.  This would induce profile $u_1$, and would result in improved result $x$.  This would contradict WSP, implying that $F(u_2) = x$.

But then, in profile $u_.$, voters with preferences $\begin{pmatrix} x \\ y \\ z\end{pmatrix}$ would want to induce profile $u_2$. 

Note that the following weights are equivalent:

$$ a - (k + m) = b + (k + m)$$

$$ b + k = 1 - a - b - k$$

$$ 1 - a - b + m = a - m$$

So then, from profile $u_.$, voters with true preferences $\begin{pmatrix} x \\ y \\ z\end{pmatrix}$ would misreport as $\begin{pmatrix} y \\ z \\ x\end{pmatrix}$ with a weight of $k$, and would misreport as $\begin{pmatrix} z \\ x \\ y \end{pmatrix}$ with a weight of $m$ (for a total coalition size of $k + m < \varepsilon$).  This would form profile $u_2$ and improved outcome $x$.  This would contradict WSP, so \textbf{Case 1.I.1.1.2} does not hold.

\textbf{Case 1.I.1.1.n+1} $ n\varepsilon \leq a - b < (n + 1)\varepsilon$

Again, as assumptions for this step, we have that:

$$ F(u_.) = y \quad (A1)$$

$$ F(u_1) = x \quad (A2) $$

Recall that:
$$ u_1 =\begin{tabular}{ c c c }
a &  b  & 1 -  a - b\\
\hline
 z  & x  & y\\ 
x  & y & z\\  
y  & z &x   
\end{tabular} =  \begin{tabular}{ c c c }
b  &   1 - a - b  & a \\
\hline
 x  & y  & z\\ 
y  & z & x\\  
z  & x & y  
\end{tabular}$$

Consider the following profile:

$$ u^{(j)} =\begin{tabular}{ c c c }
b + ($jk'$ + $jm'$) &   1 - a - b - $jk'$  & a - $jm'$\\
\hline
 x  & y  & z\\ 
y  & z & x\\  
z  & x & y  
\end{tabular} $$

(Where $k' = \frac{1 - a - 2b}{n+1}$ ; $m' = \frac{2a + b - 1}{n+1}$, and $j$ iterates from $1$ to $n$).

Suppose that $F(u^{(1)}) \neq x$.  But then, voters in profile $u^{(1)}$ with preferences $\begin{pmatrix} x \\ y \\ z\end{pmatrix}$ would misreport as $\begin{pmatrix} y \\ z \\ x\end{pmatrix}$ with a weight of $k'$, and would misreport as $\begin{pmatrix} z \\ x \\ y \end{pmatrix}$ with a weight of $m'$ (for a total coalition size of $k' + m' < \varepsilon$).  This would induce profile $u_1$, and improved result $x$.  And, this would contradict WSP, meaning that $F(u^{(1)}) = x$.

Now, assume that $F(u^{(j)}) = x$ for some general $1 \leq j < n$.  We can show that $F(u^{(j+1)}) = x$.  Suppose instead that $F(u^{(j+1)}) \neq x$.  But then, note that voters in profile $u^{(j+1)}$ with preferences $\begin{pmatrix} x \\ y \\ z\end{pmatrix}$ would misreport as $\begin{pmatrix} y \\ z \\ x\end{pmatrix}$ with a weight of $k'$, and would misreport as $\begin{pmatrix} z \\ x \\ y \end{pmatrix}$ with a weight of $m'$ (for a total coalition size of $k' + m' < \varepsilon$.  This would induce profile $u^{(j)}$, and improved result $x$.  And, this would contradict WSP.  Instead, it must be true that $F(u^{(j+1)}) = x$. 

This means that for the following profile:

$$ u^{(n)} =\begin{tabular}{ c c c }
b + ($nk'$ + $nm'$) &   1 - a - b - $nk'$  & a - $nm'$\\
\hline
 x  & y  & z\\ 
y  & z & x\\  
z  & x & y  
\end{tabular} $$

$$ F(u^{(n)}) = x$$

Then, note that:

$$ b + (nk' + nm') = b + (\frac{n(a - b)}{n+1}) = \frac{(n+1)b + na - nb}{n+1} = \frac{na + b}{n+1}$$

$$ = \frac{(n+1)a - (a - b)}{n+1} = a - (k' + m')$$

Similarly, it is also true that:

$$ 1 - a - b - nk' = b + k'$$

$$ a - nm' = 1 - a - b + m'$$

So, profile $u^{(n)}$ is exactly equivalent to the following:

$$ u^{(n)} =\begin{tabular}{ c c c }
a - ($k'$ + $m'$) &   b + $k'$ & 1 - a - b + $m'$\\
\hline
 x  & y  & z\\ 
y  & z & x\\  
z  & x & y  
\end{tabular} $$

But then, those in profile $u_.$ with preferences $\begin{pmatrix} x \\ y \\ z\end{pmatrix}$ will form a coalition of size $k' + m'$, and misreport as $\begin{pmatrix} y \\ z \\ x\end{pmatrix}$ with weight $k'$, and misreport as $\begin{pmatrix} z \\ x \\ y\end{pmatrix}$ with weight $m'$.  This will induce profile $u^{(n)}$ and improved result $x$, contradicting WSP.  This means that this case in general fails to satisfy the properties, and we have that \textbf{1.I.1.1} cannot hold.

\textbf{1.I.1.2}:  Suppose instead that $1 - a - b < b$.

For this case, we consider the following profile:

$$u_3 = \begin{tabular}{ c c c }
1 - 2b &  b  & b\\
\hline
 x  & y  & z\\ 
y  & z & x\\  
z  & x &y   
\end{tabular}$$

\textbf{1.I.1.2.1} Suppose that $F(u_3) = x \quad (A3)$.

We are still assuming that:

$$F(u_.) = y \quad (A1)$$

\textbf{Case 1.I.1.2.1.1} $0 \leq 2b + a - 1 < \varepsilon$

If $F(u_3) = x$, then voters in profile $u_.$ with preferences $\begin{pmatrix} x \\ y \\ z\end{pmatrix}$ will form a coalition of size $2b + a - 1$, and misreport as $\begin{pmatrix} z \\ x \\ y\end{pmatrix}$.  This will induce profile $u_3$ and improved result $x$, which would contradict WSP.  So, in this case, $F(u_3) \neq x$.

\textbf{Case 1.I.1.2.1.2} $\varepsilon \leq 2b + a - 1 < 2\varepsilon$

Again, for this case, we have:

$$F(u_.) = y \quad (A1)$$

And, we are assuming:

$$F(u_3) = x \quad (A3) $$

Consider the following profile:

$$u_4 = \begin{tabular}{ c c c }
a - k &  b  & 1 - a - b + k\\
\hline
 x  & y  & z\\ 
y  & z & x\\  
z  & x &y   
\end{tabular}$$

(Here, $k = \frac{2b + a - 1}{2}$.  Note that the definition of $k$ is local to this case.)

Also:

$$ a - k = 1 - 2b + k$$

$$ 1 - a - b + k = b - k$$

So:

$$u_4 = \begin{tabular}{ c c c }
1 - 2b + k &  b  & b - k\\
\hline
 x  & y  & z\\ 
y  & z & x\\  
z  & x &y   
\end{tabular}$$

Suppose that $F(u_4) \neq x$.  But then, voters in $u_4$ with preferences $\begin{pmatrix} x \\ y \\ z\end{pmatrix}$ will form a coalition of size $k$, and misreport as $\begin{pmatrix} z \\ x \\ y\end{pmatrix}$.  This will induce profile $u_3$ and improved result $x$, which would contradict WSP.  So, $F(u_4) = x$.

But then, note that voters in profile $u_.$ with preferences $\begin{pmatrix} x \\ y \\ z\end{pmatrix}$ will misreport as $\begin{pmatrix} z \\ x \\ y\end{pmatrix}$ with a weight of $k$.  This will induce profile $u_4$, and achieve improved result $x$, contradicting WSP.  So, this case fails to hold.

\textbf{Case 1.I.1.2.1.n+1} $n\varepsilon \leq 2b + a - 1 < (n+1)\varepsilon$

For this case, we have:

$$F(u_.) = y \quad (A1)$$

And, we are assuming:

$$F(u_3) = x \quad (A3) $$

$$u^{(j)} = \begin{tabular}{ c c c }
1 - 2b + j$k'$ &  b  & b - j$k'$\\
\hline
 x  & y  & z\\ 
y  & z & x\\  
z  & x &y   
\end{tabular}$$

(Here, $j$ iterates from $1$ to $n$, and $k' = \frac{2b+a - 1}{n+1}$.  Also, note that this definition of $k'$ and $u^{(j)}$ is local to this case.)

We consider:

$$u^{(1)} = \begin{tabular}{ c c c }
1 - 2b + $k'$ &  b  & b - $k'$\\
\hline
 x  & y  & z\\ 
y  & z & x\\  
z  & x &y   
\end{tabular}$$

Suppose that $F(u^{(1)}) \neq x$.  But then, voters in $u^{(1)}$ with preferences $\begin{pmatrix} x \\ y \\ z\end{pmatrix}$ will form a coalition of size $k'$, and misreport as $\begin{pmatrix} z \\ x \\ y\end{pmatrix}$.  This will induce profile $u_3$, and achieve improved outcome $x$, contradicting WSP.  So, it must be the case that $F(u^{(1)}) = x$.  

Now we assume that for some general $j$ that $F(u^{(j)}) = x$.  We can show that $F(u^{(j+1)}) = x$.  Suppose instead that $F(u^{(j+1)}) \neq x$.  But then, voters in $u^{(j+1)}$ with preferences $\begin{pmatrix} x \\ y \\ z\end{pmatrix}$ will form a coalition of size $k'$, and misreport as $\begin{pmatrix} z \\ x \\ y\end{pmatrix}$.  This will induce profile $u^{(j)}$, and will achieve improved outcome $x$.  This would contradict WSP, so, $F(u^{(j +1)}) = x$.  This is true for any $j$.

In particular, we have:
$$u^{(n)} = \begin{tabular}{ c c c }
1 - 2b + $nk'$ &  b  & b - $nk'$\\
\hline
 x  & y  & z\\ 
y  & z & x\\  
z  & x &y   
\end{tabular}$$

$$ F(u^{(n)}) = x \quad (A4)$$

Note that:

$$ 1 - 2b + nk' = 1 - 2b + \frac{n(2b+a - 1)}{n+1} =$$ 
$$ = \frac{na - 2b + 1}{n+1} = a - \frac{2b+a - 1}{n+1} = a - k'$$

Likewise:

$$ b - nk' = 1 - a -b + k'$$

So, profile $u^{(n)}$ is exactly equivalent to the following profile (with relabeled weights):

$$u^{(n)} = \begin{tabular}{ c c c }
a - $k'$ &  b  & 1 - a - b + $k'$\\
\hline
 x  & y  & z\\ 
y  & z & x\\  
z  & x &y   
\end{tabular}$$

In profile $u_.$, voters with preferences $\begin{pmatrix} x \\ y \\ z\end{pmatrix}$ will form a coalition of size $k'$, and misreport as $\begin{pmatrix} z \\ x \\ y\end{pmatrix}$.  This will induce profile $u^{(n)}$ and evaluation $(A4)$, giving improved result $x$ over $y$.  This would contradict WSP.  This means that our supposition that $F(u_3) = x$ is not possible.  So, this case also cannot hold, and concludes \textbf{1.I.1.2.1}.

\textbf{1.I.1.2.2} Now suppose that $F(u_3) = y$.

Recall that we are still in the case where $1 - a - b < b$.  Apply permutation $\sigma$ to $u_3$ (where $\sigma(x) = z, \sigma(y) = x, \sigma(z) = y$).  From A and N, this results in the following profile and evaluation:

$$ u_5 = \begin{tabular}{ c c c }
b &  b  & 1 - 2b\\
\hline
 x  & y  & z\\ 
y  & z & x\\  
z  & x &y   
\end{tabular}$$

$$ F(u_5) = x \quad (A5)$$

Note that we are still assuming:

$$ F(u_.) = y \quad (A1)$$

\textbf{Case 1.I.1.2.2.1} $0 \leq a - b < \varepsilon$

But then, voters in $u_.$ with preferences $\begin{pmatrix} x \\ y \\ z\end{pmatrix}$ will form a coalition of size $a - b < \varepsilon$, and misreport as $\begin{pmatrix} z \\ x \\ y\end{pmatrix}$.  This will induce profile $u_5$ and evaluation $(A5)$, giving improved result $x$ over $y$.  This would contradict WSP, so {Case 1.I.1.2.2.1} cannot hold.

\textbf{Case 1.I.1.2.2.n+1} $n\varepsilon \leq a - b < (n+1)\varepsilon$

Again, we have:

$$ F(u_.) = y \quad (A1)$$

And, we are assuming:

$$ F(u_5) = x \quad (A5)$$

Consider the following profile:

$$u^{(j)} = \begin{tabular}{ c c c }
b + $jk'$ &  b  & 1 - 2b - $jk'$\\
\hline
 x  & y  & z\\ 
y  & z & x\\  
z  & x &y   
\end{tabular}$$

(Here, $j$ iterates from $1$ to $n$, and $k' = \frac{a - b}{n +1}$.  Again, note that the definitions for $k', n, u^{(j)}$ are local to this section).

We consider the profile:

$$u^{(1)} = \begin{tabular}{ c c c }
b + $k'$ &  b  & 1 - 2b - $k'$\\
\hline
 x  & y  & z\\ 
y  & z & x\\  
z  & x &y   
\end{tabular}$$

Suppose that $F(u^{(1)}) \neq x$.  But then, voters in profile $u^{(1)}$ with preferences $\begin{pmatrix} x \\ y \\ z\end{pmatrix}$ will form a coalition of size $k' < \varepsilon$, and misreport as $\begin{pmatrix} z \\ x \\ y\end{pmatrix}$.  This will induce profile $u_5$ and evaluation $(A5)$, giving improved result $x$ over $y$.  This would contradict WSP, meaning that $F(u^{(1)}) = x$. 

Now, for a general $j$, assume that $F(u^{(j)}) = x$.  We can show that $F(u^{(j+1)}) = x$.  Suppose instead that $F(u^{(j+1)}) \neq x$.  But then, voters in profile $u^{(j+1)}$ with preferences $\begin{pmatrix} x \\ y \\ z\end{pmatrix}$ will form a coalition of size $k' < \varepsilon$, and misreport as $\begin{pmatrix} z \\ x \\ y\end{pmatrix}$, which would induce profile $u^{(j)}$ and improved result $x$.  This would contradict WSP.  So, in general, $F(u^{(j)}) = x$.

In particular, we have the following profile and evaluation:

$$u^{(n)} = \begin{tabular}{ c c c }
b + $nk'$ &  b  & 1 - 2b - $nk'$\\
\hline
 x  & y  & z\\ 
y  & z & x\\  
z  & x &y   
\end{tabular}$$

$$ F(u^{(n)}) = x \quad (A6) $$

Note that:

$$ b + nk' = b + \frac{n(a - b)}{n+1} = \frac{b + na}{n+1} = \frac{(n+1)a - a + b}{n+1} = a - k' $$

Similarly, it is true that:

$$ 1 - 2b - nk' = 1 - a - b + k'$$

So, profile $u^{(n)}$ is equivalent to the following relabeling of weights:

$$u^{(n)} = \begin{tabular}{ c c c }
a - $k'$ &  b  & 1 - a - b + $k'$\\
\hline
 x  & y  & z\\ 
y  & z & x\\  
z  & x &y   
\end{tabular}$$

We have that $F(u^{(n)}) = x$.  But then, in profile $u_.$, voters with preferences $\begin{pmatrix} x \\ y \\ z\end{pmatrix}$ will form a coalition of size $k'$, and misreport as $\begin{pmatrix} z \\ x \\ y\end{pmatrix}$.  This will induce profile $u^{(n)}$ and evaluation $(A6)$, giving improved result $x$ over $y$.  This would contradict WSP.

So, this case also cannot hold, meaning that for $F(u_3) \neq y$. 

\textbf{1.I.1.2.3}  Finally, suppose that  $F(u_3) = z$. 

Again, we consider profile $u_1$:

$$u_1 = \begin{tabular}{ c c c }
b  &   1 - a - b  & a \\
\hline
 x  & y  & z\\ 
y  & z & x\\  
z  & x & y  
\end{tabular}$$

Recall that: $$F(u_1) = x \quad (A2) $$

Also note that for this case, we have that $b > 1 - a - b$.

\textbf{Case 1.I.1.2.3.1} $0 \leq a - b < \varepsilon$
Here, voters in profile $u_1$ with preferences $\begin{pmatrix} z \\ x \\ y \end{pmatrix}$ will misreport as $\begin{pmatrix} x \\ y \\ z \end{pmatrix}$ with a weight of $1 - 3b$, and will misreport as $\begin{pmatrix} y \\ z \\ x \end{pmatrix}$ with a weight of $2b + a - 1$ (for a total coalition size of $a - b < \varepsilon$).  This would induce profile $u_3$, achieving improved outcome $z$.  This would contradict WSP.

\textbf{Case 1.I.1.2.3.n+1} $n\varepsilon \leq a - b < (n+1)\varepsilon$

Consider the following profile:

$$u^{(j)} = \begin{tabular}{ c c c }
1 - 2b - jk & b - jm & b + j(k + m) \\
\hline
 x  & y  & z\\ 
y  & z & x\\  
z  & x &y   
\end{tabular}$$

(Here, $k = \frac{1 - 3b}{n+1}$, $m = \frac{2b + a - 1}{n+1}$, and $j$ iterates from $1$ to $n$.  Also note that this definition of $u^{(j)}$ is local to this case). 

Suppose that $F(u^{(1)}) \neq z$.  But then, voters in profile $u^{(1)}$ with preferences $\begin{pmatrix} z \\ x \\ y \end{pmatrix}$ will misreport as $\begin{pmatrix} x \\ y \\ z \end{pmatrix}$ with a weight of $k$, and will misreport as $\begin{pmatrix} y \\ z \\ x \end{pmatrix}$ with a weight of $m$ (for a total coalition size of $k + m < \varepsilon$).  This would induce profile $u_3$, and improved result $z$.  This would contradict WSP, meaning that $F(u^{(1)}) = z$.

Now suppose for a general $j$ that $F(u^{(j)}) = z$.  We can show that $F(u^{(j+1)}) = z$ as well.  Suppose instead that $F(u^{(j+1)}) \neq z$.  But then, voters in profile $u^{(j+1)}$ with preferences $\begin{pmatrix} z \\ x \\ y \end{pmatrix}$ will misreport as $\begin{pmatrix} x \\ y \\ z \end{pmatrix}$ with a weight of $k$, and will misreport as $\begin{pmatrix} y \\ z \\ x \end{pmatrix}$ with a weight of $m$ (for a total coalition size of $k + m < \varepsilon$).  This would induce profile $u^{(j)}$, and would achieve improved result $z$, contradicting WSP.  So, $F(u^{(j+1)}) = z$.  This is true for any $j$.

In particular, we have this for:

$$u^{(n)} = \begin{tabular}{ c c c }
1 - 2b - nk & b - nm & b + n(k + m) \\
\hline
 x  & y  & z\\ 
y  & z & x\\  
z  & x &y   
\end{tabular}$$

Note that:

$$ 1 - 2b - nk = b + k$$

$$ b - nm = 1 - a - b + m$$

$$ b + n(k + m) = a - (k + m)$$

So:

$$u^{(n)} = \begin{tabular}{ c c c }
b + k & 1 - a - b + m & a - (k + m) \\
\hline
 x  & y  & z\\ 
y  & z & x\\  
z  & x &y   
\end{tabular}$$

But then, voters in profile $u_1$ with preferences $\begin{pmatrix} z \\ x \\ y \end{pmatrix}$ will misreport as $\begin{pmatrix} x \\ y \\ z \end{pmatrix}$ with a weight of $k$, and will misreport as $\begin{pmatrix} y \\ z \\ x \end{pmatrix}$ with a weight of $m$ (for a total coalition size of $k + m < \varepsilon$).  This would induce profile $u^{(n)}$, and improved result $z$.  This would contradict WSP, meaning that this case also fails.

So, in general, we see that \textbf{1.I.1} is not possible - when $y$ beats $z$ according to the Borda count, it cannot be true that $F(u_.) = y$.

\textbf{1.I.2}  Now we assume that $F$ has the following evaluation on $u_.$:

$$ F(u_.) = z \quad(A7)$$

(Or, with $u_.$ written out):

$$ F \left( \begin{tabular}{ c c c }
a &  b  & 1 -  a - b\\
\hline
 x  & y  & z\\ 
y  & z & x\\  
z  & x &y   
\end{tabular} \right) = z \quad (A7)$$

Apply permutation $\sigma$ to this profile, where $\sigma(x) = z, \sigma(y) = x, \sigma(z) = y$:

This gives us profile $u_1$, and the following evaluation:

$$ u_1 = \begin{tabular}{ c c c }
a &  b  & 1 -  a - b\\
\hline
 z  & x  & y\\ 
x  & y & z\\  
y  & z &x   
\end{tabular} = \begin{tabular}{ c c c }
b &  1 - a - b  & a\\
\hline
 x  & y  & z\\ 
y  & z & x\\  
z  & x &y   
\end{tabular}$$

$$F(u_1) = y \quad (A8)$$

\textbf{1.I.2.1} Suppose that $1 - a - b \geq b$.

\textbf{Case 1.I.2.1.1}  $0 \leq a - b < \varepsilon$

Note that voters in profile $u_.$ with preferences $\begin{pmatrix} x \\ y \\ z \end{pmatrix}$ will want to improve their outcome.  A weight of $a - (1 - a - b)$ will misreport as $\begin{pmatrix} z \\ x \\ y \end{pmatrix}$, and a weight of $(1 - a - b) - b$ will misreport as $\begin{pmatrix} y \\ z \\ x \end{pmatrix}$.  This will induce profile $u_1$ and evaluation $(A8)$, resulting in improved outcome $y$. This will have a total coalition size of $2a - 1 + b + 1 - a - 2b = a - b < \varepsilon$.  This would contradict WSP, meaning that \textbf{Case 1.I.2.1.1} cannot hold.  

\textbf{Case 1.I.2.1.n+1}  $n\varepsilon \leq a - b < (n+1)\varepsilon$

Consider the following profile:

$$ u^{(j)} =\begin{tabular}{ c c c }
a - j(k + m) &   b + jk  & 1 - a - b + jm\\
\hline
 x  & y  & z\\ 
y  & z & x\\  
z  & x & y  
\end{tabular} $$

(Here, $k = \frac{1 - a - 2b}{n+1}$ ; $m = \frac{2a + b - 1}{n+1}$.  Also note that these definitions of $u^{(j)}$ and the weights $k$ and $m$ are local to this case).

First consider $F(u^{(1)})$.  Suppose that $F(u^{(1)}) \neq z$.  But then, note that voters in profile $u_.$ with preferences $\begin{pmatrix} x \\ y \\ z \end{pmatrix}$ will want to improve their outcome.  A weight of $k$ will misreport as $\begin{pmatrix} y \\ z \\ x \end{pmatrix}$, and a weight of $m$ will misreport as $\begin{pmatrix} z \\ x \\ y \end{pmatrix}$.  This will induce profile $u^{(1)}$, giving an improved outcome over $z$. This deviation will have a total coalition size of $k + m = \frac{a - b}{n+1} < \varepsilon$.  This would contradict WSP, so $F(u^{(1)}) = z$.

Now, we assume that $F(u^{(j)}) = z$ for some general $j$.  We want to show that $F(u^{(j+1)}) = z$.  Suppose that $F(u^{(j+1)}) \neq z$.  But then, in profile $u^{(j)}$, voters with preferences $\begin{pmatrix} x \\ y \\ z \end{pmatrix}$ will misrepresent:  a weight of $k$ will misreport as $\begin{pmatrix} y \\ z \\ x \end{pmatrix}$, and a weight of $m$ will misreport as $\begin{pmatrix} z \\ x \\ y \end{pmatrix}$.  This will induce profile $u^{(j+1)}$, and result in the improved outcome over $z$.  This would contradict WSP, so, in general, we know that $F(u^{(j)}) = z$ for all $j$.

This tells us the evaluation for the following profile:

$$ u^{(n)} =\begin{tabular}{ c c c }
a - n(k + m) &   b + nk  & 1 - a - b + nm\\
\hline
 x  & y  & z\\ 
y  & z & x\\  
z  & x & y  
\end{tabular} $$

$$ F(u^{(n)}) = z$$

Note that the following weights are equivalent:

$$ a - n(k+m) = \frac{(n+1)a - na + nb}{n+1} = \frac{a+nb}{n+1}=b + (k+m)$$

$$ b + nk = \frac{(n+1)b + n - na - 2nb}{n+1} = \frac{b - nb + n - na}{n+1} = 1 - a - b - k$$

$$ 1 - a - b + nm = a - m$$

So, equivalently for $u^{(n)}$, we have:

$$ u^{(n)} =\begin{tabular}{ c c c }
b + (k + m) &   1 - a - b - k  & a - m\\
\hline
 x  & y  & z\\ 
y  & z & x\\  
z  & x & y  
\end{tabular} $$

We recall that $F(u_1) = y$.  Voters in profile $u^{(n)}$ will misreport their preferences as $\begin{pmatrix} y \\ z \\ x \end{pmatrix}$ with a weight of $k$, and a weight of $m$ will misreport as $\begin{pmatrix} z \\ x \\ y \end{pmatrix}$.  This will induce profile $u_1$ and evaluation $(A8)$, yielding improved outcome $y$.  So, in general, this case cannot hold, and this concludes \textbf{1.I.2.1}.

\textbf{1.I.2.2} Now we suppose that $b > 1 - a - b$.

We apply permutation $\nu$ to profile $u_.$ (where $\nu(x) = y, \nu(y) = z, \nu(z) = x$).  By A, N, and $(A7)$, this gives us:

$$u_6 = \begin{tabular}{ c c c }
1 - a - b &  a  & b\\
\hline
 x  & y  & z\\ 
y  & z & x\\  
z  & x &y   
\end{tabular} $$

$$ F(u_6) = x \quad (A9)$$

\textbf{Case 1.I.2.2.1} $0 \leq 2a + b - 1 < \varepsilon$

Voters in profile $u_.$ with preferences $\begin{pmatrix} x \\ y \\ z \end{pmatrix}$ will misreport:  a weight of $(a - b)$ will misreport as $\begin{pmatrix} y \\ z \\ x \end{pmatrix}$, and a weight of $(2b + a - 1)$ will misreport as $\begin{pmatrix} z \\ x \\ y \end{pmatrix}$.  The resulting coalition size is $2a + b - 1<\varepsilon$.  This will induce profile $u_6$ and improved outcome $x$, which would contradict WSP.  So, \textbf{Case 1.I.2.2.1} cannot hold.

\textbf{Case 1.I.2.2.n+1} $n\varepsilon \leq 2a + b - 1 < (n+1)\varepsilon$

Now, consider the following profile:

$$ u^{(j)} =\begin{tabular}{ c c c }
a - j(k + m) &   b + jk  & 1 - a - b + jm\\
\hline
 x  & y  & z\\ 
y  & z & x\\  
z  & x & y  
\end{tabular} $$

(Here, $k = \frac{a - b}{n+1}$, $m = \frac{2b + a - 1}{n+1}$, and $j$ iterates from $1$ to $n$.  Note that the definition of $u^{(j)}$ is local to this case).

Consider $F(u^{(1)})$.  Suppose that $F(u^{(1)}) \neq z$.  But then, voters in profile $u_.$ with preferences $\begin{pmatrix} x \\ y \\ z \end{pmatrix}$ will misreport:  a weight of $k$ will misreport as $\begin{pmatrix} y \\ z \\ x \end{pmatrix}$, and a weight of $m$ will misreport as $\begin{pmatrix} z \\ x \\ y \end{pmatrix}$.  The resulting coalition size is $k + m = \frac{2a + b - 1}{n+1}<\varepsilon$.  This would induce profile $u^{(1)}$, and give an improved outcome over $z$.  This would contradict WSP, so, we know that $F(u^{(1)}) = z$.

Suppose for some $j$ that $F(u^{(j)}) = z$.  We can show that $F(u^{(j+1)}) = z$.  Suppose instead that $F(u^{(j+1)} \neq z$.  But then, voters in $F(u^{(j)})$ with preferences $\begin{pmatrix} x \\ y \\ z \end{pmatrix}$ will misreport:  a weight of $k$ will misreport as $\begin{pmatrix} y \\ z \\ x \end{pmatrix}$, and a weight of $m$ will misreport as $\begin{pmatrix} z \\ x \\ y \end{pmatrix}$. Again, the total coalition size is $k + m <\varepsilon$.  This would induce profile $u^{(j+1)}$ and an improved outcome over $z$ for those who deviated.  So, this would contradict WSP, meaning that $F(u^{(j+1)}) = z$.

This means that:

$$ u^{(n)} =\begin{tabular}{ c c c }
a - n(k + m) &   b + nk  & 1 - a - b + nm\\
\hline
 x  & y  & z\\ 
y  & z & x\\  
z  & x & y  
\end{tabular} $$

$$ F(u^{(n)}) = z$$

Note that:

$$ a - n(k + m) = 1 - a - b + (k+m)$$

$$ b + nk = \frac{b(n+1) + n(a - b)}{n+1} = \frac{b + na}{n+1} = \frac{(n+1)a - (a - b)}{n+1}=  a - k$$

$$ 1 - a - b + nm =  \frac{n + 1 - na - a - nb - b + 2nb + na - n}{n+1}=$$ 

$$ = \frac{1 - a - 2b + b + nb}{n+1} = \frac{(n+1)b - (2b + a - 1)}{n+1}=  b - m$$

So:

$$ u^{(n)} =\begin{tabular}{ c c c }
1 - a - b + (k + m) &   a - k  & b - m \\
\hline
 x  & y  & z\\ 
y  & z & x\\  
z  & x & y  
\end{tabular} $$

So then, voters in profile $u^{(n)}$ with preferences $\begin{pmatrix} x \\ y \\ z \end{pmatrix}$ will misreport:  a weight of $k$ will misreport as $\begin{pmatrix} y \\ z \\ x \end{pmatrix}$, and a weight of $m$ will misreport as $\begin{pmatrix} z \\ x \\ y \end{pmatrix}$.  This will induce profile $u_6$, and improved result $x$.  This contradicts WSP, meaning that \textbf{Case 1.I.2.2.n+1} cannot hold.

This concludes \textbf{1.I.2}, and means that \textbf{Case I} of \textbf{Step 1} is impossible.  

So, we move on to \textbf{Case II} of Step 1.

\newpage

\textbf{Step 1, Case II:} Now suppose that by the Borda Count, $z$ beats $y$. 

So, we have the following inequalities:

$$ a + 1 - b > 2 - 2a -b > a + 2b$$

$$ a + b < \frac{2}{3} $$

$$ a > \frac{1}{3} > b$$

Again, we are considering the following profile:

$$u_. =  \begin{tabular}{ c c c }
a &  b  & 1 -  a - b\\
\hline
 x  & y  & z\\ 
y  & z & x\\  
z  & x &y   
\end{tabular}$$

\textbf{1.II.1} Suppose that $F$ has the following evaluation on $u_.$:

$$F(u_.) = z \quad (A10)$$

We apply permutation $\sigma$ to $u_.$ (where $\sigma(x) = z, \sigma(y) = x, \sigma(z) = y$),  resulting in profile $u_1$ from before, and the following evaluation:

$$ u_1 =\begin{tabular}{ c c c }
a &  b  & 1 -  a - b\\
\hline
 z  & x  & y\\ 
x  & y & z\\  
y  & z &x   
\end{tabular} = \begin{tabular}{ c c c }
b &  1 - a - b  & a\\
\hline
 x  & y  & z\\ 
y  & z & x\\  
z  & x &y   
\end{tabular} $$

$$ F(u_1) = y \quad (A11)$$

\textbf{1.II.1.1} Suppose that $a \geq 1 - a - b$.

And, note that $1 - a - b > b$. 

\textbf{Case 1.II.1.1.1} $0 \leq a - b < \varepsilon$

Voters in profile $u_.$ with preferences $\begin{pmatrix} x \\ y \\ z \end{pmatrix}$ misreport to improve their outcome, where a weight of $1 - a - 2b$ will misreport as $\begin{pmatrix} y \\ z \\ x \end{pmatrix}$, and a weight of $2a - 1 + b$ will misreport as $\begin{pmatrix} z \\ x \\ y \end{pmatrix}$.   This deviation has a coalition size of $a - b < \varepsilon$.  This will induce profile $u_1$ and evaluation $(A11)$, for improved result $y$.

\textbf{Case 1.II.1.1.n+1} $ n\varepsilon \leq a - b < (n + 1)\varepsilon$

Again, by assumption, we have that:

$$ F(u_.) = z \quad (A10)$$

And, we are assuming:

$$ F(u_1) = y \quad (A11) $$

Consider the following profile:

$$ u^{(j)} =\begin{tabular}{ c c c }
a - (jk + jm) &   b + jk  & 1 - a - b + jm\\
\hline
 x  & y  & z\\ 
y  & z & x\\  
z  & x & y  
\end{tabular} $$

(Here, $k = \frac{1 - a - 2b}{n+1}; m = \frac{2a + b - 1}{n+1}$, and $j$ iterates from $1$ to $n$.  Also note that this definition of $u^{(j)}$ is local to this case).

Suppose that $F(u^{(1)}) \neq z$.  But then, voters in profile $u_.$ with preferences $\begin{pmatrix} x \\ y \\ z\end{pmatrix}$ would misreport as $\begin{pmatrix} y \\ z \\ x\end{pmatrix}$ with a weight of $k$, and would misreport as $\begin{pmatrix} z \\ x \\ y \end{pmatrix}$ with a weight of $m$ (for a total coalition size of $k + m < \varepsilon$).  This would induce profile $u^{(1)}$, and an improved result over $z$.  And, this would contradict WSP, meaning that $F(u^{(1)}) = z$.

Now, assume that $F(u^{(j)}) = z$ for some general $1 \leq j < n$.  We can show that $F(u^{(j+1)}) = z$.  Suppose instead that $F(u^{(j+1)}) \neq z$.  But then, note that voters in profile $u^{(j)}$ with preferences $\begin{pmatrix} x \\ y \\ z\end{pmatrix}$ would misreport as $\begin{pmatrix} y \\ z \\ x\end{pmatrix}$ with a weight of $k$, and would misreport as $\begin{pmatrix} z \\ x \\ y \end{pmatrix}$ with a weight of $m$ (for a total coalition size of $k + m < \varepsilon$).  This would induce profile $u^{(j+1)}$, and an improved result over $z$, which would contradict WSP.  Instead, it must be true that $F(u^{(j+1)}) = z$. 

This means that for the following profile:

$$ u^{(n)} =\begin{tabular}{ c c c }
a - (nk + nm) &   b + nk  & 1 - a - b + nm\\
\hline
 x  & y  & z\\ 
y  & z & x\\  
z  & x & y  
\end{tabular} $$

$$ F(u^{(n)}) = z$$

Then, note that:

$$ a - (nk + nm) = b + (k + m)$$

Similarly, it is also true that:

$$ b + nk = 1 - a - b - k$$

$$ 1 - a - b + nm = a - m$$

So, profile $u^{(n)}$ is exactly equivalent to the following:

$$ u^{(n)} =\begin{tabular}{ c c c }
b + (k + m) & 1 - a - b - k & a - m\\
\hline
 x  & y  & z\\ 
y  & z & x\\  
z  & x & y  
\end{tabular} $$

But then, those in profile $u^{(n)}$ with preferences $\begin{pmatrix} x \\ y \\ z\end{pmatrix}$ will form a coalition of size $k + m$, and misreport as $\begin{pmatrix} y \\ z \\ x\end{pmatrix}$ with weight $k$, and misreport as $\begin{pmatrix} z \\ x \\ y\end{pmatrix}$ with weight $m$.  This will induce profile $u_1$ and evaluation $(A11)$, yielding improved result $y$, contradicting WSP.  This means that this case in general fails to satisfy the properties, and we have that \textbf{1.II.1.1} cannot hold.

\textbf{1.II.1.2} Now assume instead that $a < 1 - a - b$.

Consider the following profile:

$$ u_7 = \begin{tabular}{ c c c }
a &   1 - 2a  & a\\
\hline
 x  & y  & z\\ 
y  & z & x\\  
z  & x & y  
\end{tabular} $$

\textbf{1.II.1.2.1} Suppose that $F(u_7) = x \quad (A12)$.

Also, recall that we are assuming:

$$ F(u_.) = z \quad (A10) $$

\textbf{Case 1.II.1.2.1.1} $0 \leq 1 - 2a - b < \varepsilon$

Note that voters in profile $u_7$ with preferences $\begin{pmatrix} y \\ z \\ x\end{pmatrix}$ will misreport their preferences as $\begin{pmatrix} z \\ x \\ y \end{pmatrix}$ with a weight of $1 - 2a - b < \varepsilon$.  This will induce profile $u_.$, and achieve improved outcome $z$.  This would contradict WSP, so \textbf{Case 1.II.1.2.1.1} fails.

\textbf{Case 1.II.1.2.1.n+1} $n\varepsilon \leq 1 - 2a - b < (n+1)\varepsilon$

Now, consider the following profile:

$$ u^{(j)} = \begin{tabular}{ c c c }
a &   1 - 2a - jk  & a + jk\\
\hline
 x  & y  & z\\ 
y  & z & x\\  
z  & x & y  
\end{tabular}$$

(Where $k = \frac{1 - 2a - b}{n+1}$, and $j$ iterates from $1$ to $n$.  Also, note that this definition of $u^{(j)}$ is local to this case).

Suppose that $F(u^{(1)}) \neq x$.  But then, voters in profile $u_7$ with preferences $\begin{pmatrix} y \\ z \\ x\end{pmatrix}$ will misreport their preferences as $\begin{pmatrix} z \\ x \\ y \end{pmatrix}$ with a weight of $k < \varepsilon$.  This would induce profile $u^{(1)}$ and an improved result over $x$, contradicting WSP.  So,  $F(u^{(1)}) = x$.

Now suppose that $F(u^{(j)}) = x$ for some general $j$.  We can show that $F(u^{(j+1)}) = x$.  Suppose instead that $F(u^{(j+1)}) \neq x$.  But then, voters in profile $u^{(j)}$ with preferences $\begin{pmatrix} y \\ z \\ x\end{pmatrix}$ will misreport their preferences as $\begin{pmatrix} z \\ x \\ y \end{pmatrix}$ with a weight of $k < \varepsilon$.  This would induce profile $u^{(j+1)}$ and an improved result over $x$, contradicting WSP.  This means that $F(u^{(j+1)}) = x$ for any general $j$.

Note that in particular, we have this for:

$$ u^{(n)} = \begin{tabular}{ c c c }
a &   1 - 2a - nk  & a + nk\\
\hline
 x  & y  & z\\ 
y  & z & x\\  
z  & x & y  
\end{tabular} = \begin{tabular}{ c c c }
a &   b + k  & 1 - a - b - k\\
\hline
 x  & y  & z\\ 
y  & z & x\\  
z  & x & y  
\end{tabular}$$

But then, voters in profile $u^{(n)}$ with preferences $\begin{pmatrix} y \\ z \\ x\end{pmatrix}$ will misreport their preferences as $\begin{pmatrix} z \\ x \\ y \end{pmatrix}$ with a weight of $k < \varepsilon$.  This will induce profile $u_.$, and achieve improved outcome $z$ over $x$.  This contradicts WSP, so in general, \textbf{Case 1.II.1.2.1} cannot hold.

\textbf{1.II.1.2.2} Now suppose that $F(u_7) = y \quad (A13)$.

Again, recall that we are assuming:

$$ F(u_.) = z \quad (A10) $$

Apply permutation $\sigma$ to $u_7$, where $\sigma(x) = z, \sigma(y) = x, \sigma(z) = y$.  By properties A and N, this gives us:

$$ u_8 = \begin{tabular}{ c c c }
1 - 2a &   a  & a\\
\hline
 x  & y  & z\\ 
y  & z & x\\  
z  & x & y  
\end{tabular} $$

$$ F(u_8) = x \quad (A14)$$

\textbf{Case 1.II.1.2.2.1} $0 \leq 3a - 1 < \varepsilon$

Note that voters in profile $u_7$ with preferences $\begin{pmatrix} x \\ y \\ z \end{pmatrix}$ will misreport their preferences as $\begin{pmatrix} y \\ z \\ x \end{pmatrix}$ with a coalition size of $3a - 1 < \varepsilon$.  This would induce profile $u_8$, and would achieve improved result $x$, contradicting WSP.  So, \textbf{Case 1.II.1.2.2.1} cannot hold.

\textbf{Case 1.II.1.2.2.n+1} $n\varepsilon \leq 3a - 1 < (n+1)\varepsilon$

Now, consider the following profile:

$$ u^{(j)} = \begin{tabular}{ c c c }
1 - 2a + jk &   a - jk  & a\\
\hline
 x  & y  & z\\ 
y  & z & x\\  
z  & x & y  
\end{tabular}$$

(Here, $k = \frac{3a - 1}{n+1}$, and $j$ iterates from $1$ to $n$.  Again, we note that this definition of $u^{(j)}$ is local to this case).

Suppose that $F(u^{(1)}) \neq x$.  But then, voters in profile $u^{(1)}$ with preferences $\begin{pmatrix} x \\ y \\ z \end{pmatrix}$ will misreport their preferences as $\begin{pmatrix} y \\ z \\ x \end{pmatrix}$ with a coalition size of $k < \varepsilon$.  This would induce profile $u_8$, and would achieve improved result $x$, contradicting WSP.  So, $F(u^{(1)}) = x$.

Now, we assume that $F(u^{(j)}) = x$ for some general $j$.  We can show that  $F(u^{(j+1)}) = x$ as well.  Suppose instead that $F(u^{(j+1)}) \neq x$.  But then, note that voters in profile $u^{(j+1)}$ with preferences $\begin{pmatrix} x \\ y \\ z \end{pmatrix}$ will misreport their preferences as $\begin{pmatrix} y \\ z \\ x \end{pmatrix}$ with a coalition size of $k < \varepsilon$.  This misrepresentation would induce profile $u^{(j)}$, and would lead to improved result $x$.  This would contradict WSP, meaning that $F(u^{(j+1)}) = x$ for any $j$.

In particular, this is true for the following profile:

$$ u^{(n)} = \begin{tabular}{ c c c }
1 - 2a + nk &   a - nk  & a\\
\hline
 x  & y  & z\\ 
y  & z & x\\  
z  & x & y  
\end{tabular} =  \begin{tabular}{ c c c }
a - k &   1 - 2a + k  & a\\
\hline
 x  & y  & z\\ 
y  & z & x\\  
z  & x & y  
\end{tabular}$$

But then, voters in profile $u_7$ with preferences $\begin{pmatrix} x \\ y \\ z \end{pmatrix}$ will misreport their preferences as $\begin{pmatrix} y \\ z \\ x \end{pmatrix}$ with a coalition size of $k < \varepsilon$, inducing profile $u^{(n)}$ and achieving improved result $x$ over $y$.  This would contradict WSP.  So, \textbf{1.II.1.2.2} also cannot hold.

\textbf{1.II.1.2.3}  Now we suppose that $F(u_7) = z \quad (A15)$.

Again, by assumption for \textbf{1.II.1}, we have:

$$ F(u_.) = z \quad (A10) $$

Apply permutation $\sigma$ to $u_7$. 

This gives us profile $u_8$ and the following evaluation (from properties A and N):

$$ u_8 = \begin{tabular}{ c c c }
1 - 2a &   a  & a\\
\hline
 x  & y  & z\\ 
y  & z & x\\  
z  & x & y  
\end{tabular} $$

$$ F(u_8) = y \quad (A16)$$

\textbf{Case 1.II.1.2.3.1} $0 \leq 3a - 1 < \varepsilon$

Note that voters in profile $u_7$ with preferences $\begin{pmatrix} x \\ y \\ z \end{pmatrix}$ will misreport their preferences as $\begin{pmatrix} y \\ z \\ x \end{pmatrix}$ with a coalition size of $3a - 1 < \varepsilon$.  This would induce profile $u_8$, and would achieve improved result $y$ over $z$ contradicting WSP.  So, \textbf{Case 1.II.1.2.3.1} cannot hold.

\textbf{Case 1.II.1.2.3.n+1} $n\varepsilon \leq 3a - 1 < (n+1)\varepsilon$

Now, consider the following profile:

$$ u^{(j)} = \begin{tabular}{ c c c }
a - jk &   1 - 2a + jk  & a\\
\hline
 x  & y  & z\\ 
y  & z & x\\  
z  & x & y  
\end{tabular}$$

(Here, $k = \frac{3a - 1}{n+1}$, and $j$ iterates from $1$ to $n$.  Again, we note that this definition of $u^{(j)}$ is local to this case).

Suppose that $F(u^{(1)}) \neq z$.  But then, voters in profile $u_7$ with preferences 
$\begin{pmatrix} x \\ y \\ z \end{pmatrix}$ will misreport their preferences as $\begin{pmatrix} y \\ z \\ x \end{pmatrix}$ with a coalition size of $k < \varepsilon$.  This would induce profile $u^{(1)}$, and would achieve an improved result over $z$ contradicting WSP.  So, $F(u^{(1)}) = z$.

Now, we suppose for a general $j$ that $F(u^{(j)}) = z$.  We can then show that $F(u^{(j+1)}) = z$.  Suppose instead that $F(u^{(j+1)}) \neq z$.  But then, voters in profile $u^{(j)}$ with preferences 
$\begin{pmatrix} x \\ y \\ z \end{pmatrix}$ will misreport their preferences as $\begin{pmatrix} y \\ z \\ x \end{pmatrix}$ with a coalition size of $k < \varepsilon$.  This would induce profile $u^{(j+1)}$, and would achieve an improved outcome over $z$.  This would contradict WSP, so, for any general $j$, $F(u^{(j+1)}) = z$.

In particular, note that this is true for the following profile:

$$ u^{(n)} = \begin{tabular}{ c c c }
a - nk &   1 - 2a + nk  & a\\
\hline
 x  & y  & z\\ 
y  & z & x\\  
z  & x & y  
\end{tabular} = \begin{tabular}{ c c c }
1 - 2a + k &   a - k  & a\\
\hline
 x  & y  & z\\ 
y  & z & x\\  
z  & x & y  
\end{tabular}$$

But then, note that voters in profile $u^{(n)}$ with preferences 
$\begin{pmatrix} x \\ y \\ z \end{pmatrix}$ will misreport their preferences as $\begin{pmatrix} y \\ z \\ x \end{pmatrix}$ with a coalition size of $k < \varepsilon$. This will induce profile $u_8$ and evaluation $(A16)$, giving improved result $y$ over $z$.  This contradicts WSP.

So, these cases are exhaustive for \textbf{1.II.1}, and we conclude that when $z$ beats $y$ according to the Borda count, then $F(u_.) \neq z$.  We move on to \textbf{1.II.2}

\textbf{1.II.2} Now suppose that $F$ has the following evaluation on $u_.$:

$$ u_. = \begin{tabular}{ c c c }
a &  b  & 1 -  a - b\\
\hline
 x  & y  & z\\ 
y  & z & x\\  
z  & x &y   
\end{tabular} $$

$$F(u_.) = y \quad (A17)$$

Apply permutation $\sigma$ to profile $u_.$ and evaluation $(A17)$ where $\sigma(x) = z, \sigma(y) = x, \sigma(z) = y$.  This gives us:

$$ u_1 =\begin{tabular}{ c c c }
a &  b  & 1 -  a - b\\
\hline
 z  & x  & y\\ 
x  & y & z\\  
y  & z &x   
\end{tabular} = \begin{tabular}{ c c c }
b &  1 - a - b  & a\\
\hline
 x  & y  & z\\ 
y  & z & x\\  
z  & x &y   
\end{tabular} $$

By A and N, we have:

$$ F(u_1) = x \quad (A18)$$

\textbf{1.II.2.1} Suppose that $a \geq 1 - a - b$

Again, note that $1 - a - b > b$.

\textbf{Case 1.II.2.1.1} $0 \leq a - b < \varepsilon$

In this case, voters in profile $u_.$ with preferences $\begin{pmatrix} x \\ y \\ z \end{pmatrix}$ will misreport as $\begin{pmatrix} z \\ x \\ y \end{pmatrix}$  with a weight of $a - (1 - a - b)$ and will misreport as $\begin{pmatrix} y \\ z \\ x \end{pmatrix}$  with a weight of $1 - a - 2b$, for a total coalition size of $a - b < \varepsilon$.  This will induce profile $u_1$ and improved result $x$.  This would contradict WSP, meaning that this case cannot hold.

\textbf{Case 1.II.2.1.n+1} $n\varepsilon \leq a - b < (n+1)\varepsilon$

Now, consider the following profile:

$$ u^{(j)} =\begin{tabular}{ c c c }
b + jk + jm &   1 - a - b - jk  & a - jm \\
\hline
 x  & y  & z\\ 
y  & z & x\\  
z  & x & y  
\end{tabular} $$

(Where $k = \frac{1 - a - 2b}{n+1}$ ; $m = \frac{2a + b - 1}{n+1}$, and $j$ iterates from $1$ to $n$.  Also, note that the definition of $u^{(j)}$ is local to this case).

Suppose that $F(u^{(1)}) \neq x$.  But then, voters in profile $u^{(1)}$ with preferences $\begin{pmatrix} x \\ y \\ z \end{pmatrix}$ will misreport as $\begin{pmatrix} y \\ z \\ x \end{pmatrix}$  with a weight of $k$ and will misreport as $\begin{pmatrix} z \\ x \\ y \end{pmatrix}$  with a weight of $m$.  This would induce profile $u_1$ and evaluation $(A18)$, for improved outcome $x$.  This would contradict WSP, so $F(u^{(1)}) = x$.

Now assume for some general $j$ that $F(u^{(j)}) = x$.  We can show that $F(u^{(j+1)}) = x$.  Suppose instead that $F(u^{(j+1)}) \neq x$.  But then, voters in profile $u^{(j+1)}$ with preferences $\begin{pmatrix} x \\ y \\ z \end{pmatrix}$ will misreport as $\begin{pmatrix} y \\ z \\ x \end{pmatrix}$  with a weight of $k$ and will misreport as $\begin{pmatrix} z \\ x \\ y \end{pmatrix}$  with a weight of $m$.  This would induce profile $u^{(j)}$ and an improved outcome $x$.  This would contradict WSP, meaning that for any general $j$, $F(u^{(j+1)}) = x$.

In particular, we have this for:

$$ u^{(n)} =\begin{tabular}{ c c c }
b + nk + nm &   1 - a - b - nk  & a - nm \\
\hline
 x  & y  & z\\ 
y  & z & x\\  
z  & x & y  
\end{tabular} $$

Note that:

$$ b + n(k + m) = a - (k + m)$$

$$ 1 - a - b - nk = b + k $$

$$ a - nm = 1 - a - b + m$$

So:

$$ u^{(n)} =\begin{tabular}{ c c c }
a - (k + m) &   b + k  & 1 - a - b + m \\
\hline
 x  & y  & z\\ 
y  & z & x\\  
z  & x & y  
\end{tabular} $$

But then, note that voters in profile $u_.$ with preferences $\begin{pmatrix} x \\ y \\ z \end{pmatrix}$ will misreport as $\begin{pmatrix} y \\ z \\ x \end{pmatrix}$  with a weight of $k$ and will misreport as $\begin{pmatrix} z \\ x \\ y \end{pmatrix}$  with a weight of $m$, for a total coalition size of $k + m < \varepsilon$.  This would induce profile $u^{(n)}$ and an improved outcome $x$.  This would contradict WSP.  So, in general, we see that \textbf{1.II.2.1} fails to hold.

\textbf{1.II.2.2} Suppose that $a < 1 - a - b$.

Consider the following profile:

$$ u_9 = \begin{tabular}{ c c c }
a & 1 - 2a & a\\
\hline
 x  & y  & z\\ 
y  & z & x\\  
z  & x & y  
\end{tabular}$$

Note that we are still assuming:

$$ u_. = \begin{tabular}{ c c c }
a &  b  & 1 -  a - b\\
\hline
 x  & y  & z\\ 
y  & z & x\\  
z  & x &y   
\end{tabular} $$

$$F(u_.) = y \quad (A17)$$

\textbf{1.II.2.2.1} Suppose that:

$$ F(u_9) \neq y \quad (A19)$$

\textbf{Case 1.II.2.2.1.1} $0 \leq 1 - 2a - b < \varepsilon$

But then, voters in profile $u_9$ with preferences $\begin{pmatrix} y \\ z \\ x \end{pmatrix}$ will misreport as $\begin{pmatrix} z \\ x \\ y \end{pmatrix}$ with a coalition size of $1 - 2a - b$.  This will induce profile $u_.$ and evaluation $(A17)$, for improved outcome $y$.  This would contradict WSP, so this case fails to hold.

\textbf{Case 1.II.2.2.1.n+1} $n\varepsilon \leq 1 - 2a - b < (n+1)\varepsilon$

Consider the following profile:

$$ u^{(j)} = \begin{tabular}{ c c c }
a &   1 - 2a - jk  & a + jk\\
\hline
 x  & y  & z\\ 
y  & z & x\\  
z  & x & y  
\end{tabular}$$

(Here, $k = \frac{1 - 2a -b}{n+1}$, and $j$ iterates from $1$ to $n$.  Again, we note that this definition of $u^{(j)}$ is local to this case).

Suppose that $F(u^{(1)}) = y$.  But then, voters in profile $u_9$ with preferences $\begin{pmatrix} y \\ z \\ x \end{pmatrix}$ will misreport as $\begin{pmatrix} z \\ x \\ y \end{pmatrix}$ with a coalition size of $k < \varepsilon$.  This will induce profile $u^{(1)}$, for improved outcome $y$.  This would contradict WSP, meaning that $F(u^{(1)}) \neq y$.

Now, we assume that $F(u^{(j)}) \neq y$ for some general $j$.  We can show that $F(u^{(j+1)}) \neq y$ as well.  Suppose instead that $F(u^{(j+1)}) = y$.  But then, voters in profile $u^{(j)}$ with preferences $\begin{pmatrix} y \\ z \\ x \end{pmatrix}$ will misreport as $\begin{pmatrix} z \\ x \\ y \end{pmatrix}$ with a coalition size of $k < \varepsilon$.  This will induce profile $u^{(j+1)}$, for improved outcome $y$.  This would contradict WSP, meaning that $F(u^{(j+1)}) \neq y$.

This is true for the following profile:

$$ u^{(n)} = \begin{tabular}{ c c c }
a &   1 - 2a - nk  & a + nk \\
\hline
 x  & y  & z\\ 
y  & z & x\\  
z  & x & y  
\end{tabular} = \begin{tabular}{ c c c }
a &   b + k  & 1 - a - b - k\\
\hline
 x  & y  & z\\ 
y  & z & x\\  
z  & x & y  
\end{tabular} $$

But then, note that voters in profile $u^{(n)}$ with preferences $\begin{pmatrix} y \\ z \\ x \end{pmatrix}$ will misreport as $\begin{pmatrix} z \\ x \\ y \end{pmatrix}$ with a coalition size of $k < \varepsilon$.  This would induce profile $u_.$ and evaluation $(A17)$, for improved result $y$.  This would contradict WSP, meaning that this case cannot hold.

\textbf{1.II.2.2.2}  So instead, it must be true that:

$$ F(u_9) = y \quad (A20)$$

Apply permutation $\sigma$ to profile $u_9$ (where $\sigma(x) = z, \sigma(y) = x, \sigma(z) = y$), which, by N and A, gives us:

$$ u_{10} = \begin{tabular}{ c c c }
1 - 2a & a & a\\
\hline
 x  & y  & z\\ 
y  & z & x\\  
z  & x & y  
\end{tabular} $$

$$ F(u_{10}) = x \quad (A21)$$

\textbf{Case 1.II.2.2.2.1} $0 \leq 3a - 1 < \varepsilon$

Note that voters in profile $u_9$ with preferences $\begin{pmatrix} x \\ y \\ z \end{pmatrix}$ will misreport as $\begin{pmatrix} y \\ z \\ x \end{pmatrix}$ with a coalition weight of $3a - 1 < \varepsilon$.  This will induce profile $u_{10}$, and achieve improved outcome $x$.  This would contradict WSP, meaning that \textbf{Case 1.II.2.2.2.1} cannot hold.

\textbf{Case 1.II.2.2.2.n+1} $n\varepsilon \leq 3a - 1 < (n+1)\varepsilon$

Now, consider the following profile:

$$ u^{(j)} = \begin{tabular}{ c c c }
1 - 2a + jk &   a - jk & a\\
\hline
 x  & y  & z\\ 
y  & z & x\\  
z  & x & y  
\end{tabular}$$

(Here, $k = \frac{3a - 1}{n+1}$, and $j$ iterates from $1$ to $n$.  Also note that the definition of $u^{(j)}$ is local to this case).

Suppose that $F(u^{(1)}) \neq x$.  But then, voters in profile $u^{(1)}$ with preferences $\begin{pmatrix} x \\ y \\ z \end{pmatrix}$ will misreport as $\begin{pmatrix} y \\ z \\ x \end{pmatrix}$ with a coalition weight of $k < \varepsilon$.  This will induce profile $u_{10}$, and achieve improved outcome $x$.  This would contradict WSP, meaning that $F(u^{(1)}) = x$.

Now, assume that for some general $j$ that $F(u^{(j)}) = x$.  We can show that $F(u^{(j+1)}) = x$.  Suppose instead that $F(u^{(j+1)}) \neq x$.  But then, voters in profile $u^{(j+1)}$ with preferences $\begin{pmatrix} x \\ y \\ z \end{pmatrix}$ would misreport as $\begin{pmatrix} y \\ z \\ x \end{pmatrix}$ with a coalition weight of $k < \varepsilon$.  This would induce profile $u^{(j)}$, and result in improved outcome $x$, which would contradict WSP.  So, $F(u^{(j+1)}) = x$.

In particular, this is true for the following profile:

$$ u^{(n)} = \begin{tabular}{ c c c }
1 - 2a + nk &   a - nk & a\\
\hline
 x  & y  & z\\ 
y  & z & x\\  
z  & x & y  
\end{tabular} = \begin{tabular}{ c c c }
a - k &   1 - 2a + k & a\\
\hline
 x  & y  & z\\ 
y  & z & x\\  
z  & x & y  
\end{tabular}$$

But then, voters in profile $u_{9}$ with preferences $\begin{pmatrix} x \\ y \\ z \end{pmatrix}$ will misreport as $\begin{pmatrix} y \\ z \\ x \end{pmatrix}$ with a coalition weight of $k < \varepsilon$.  This will induce profile $u^{(n)}$, giving improved result $x$, and would violate WSP.

But this means that \textbf{Case II} is impossible.  We have shown that there is no voting rule that differs from the Borda Count which satisfies the properties P, A, N, and WSP on the Condorcet Cycle domain.  This concludes \textbf{Step 1}.

\newpage
\section*{Step 2}

Now, we want to show that there is no voting rule that satisfies the axioms P, A, N, and WSP on any expansion of the Condorcet cycle domain; by symmetry, it suffices to show that there is no voting rule on:

$$U^{*} = \Bigg\{ \begin{tabular}{ c c c c}
 x  & y & z & x\\ 
y  & z & x & z\\  
z  & x & y & y    
\end{tabular} \Bigg\} $$

Proceed by contradiction.  Assume that $F$ is a voting rule that satisfies the properties.  Without loss of generality, we assume $\frac{2}{3} > \varepsilon >0.$  

We know that on the Condorcet domain, the only voting rule that satisfies P, A, N, and WSP is the Borda Count.  So, this tells us how $F$ must behave on the following profile:

$$ u_1 = \begin{tabular}{ c c c }
(1/3) - d &  (1/3) + 2d  & (1/3) - d\\
\hline
 x  & y  & z\\ 
y  & z & x\\  
z  & x &y   
\end{tabular}$$

$$ F(u_1) = y \quad (E1)$$

Let $0 < d < \frac{1}{12}$.  We define $d$ based on $\varepsilon$:

$$d = \frac{\varepsilon}{8} < \frac{2}{24} = \frac{1}{12}$$

\smallskip

For this step, we will consider the following profile:

$$u_2 = \begin{tabular}{ c c c }
(1/3) - d &  (1/3) + 2d  & (1/3) - d\\
\hline
 x  & y  & x\\ 
z  & z & z\\  
y  & x &y   
\end{tabular} \quad $$

We will show that $F(u_2)$ cannot be defined in a way that satisfies the properties P, A, N, and WSP.

\textbf{2.I}  Suppose that the profile $u_2$ has the following evaluation.
$$F(u_2) =  x \quad (E2)$$

We will show that this cannot be true for any $d$ and $\varepsilon$, indexing by the interval $n\varepsilon \leq \frac{2}{3} - 2d < (n + 1)\varepsilon$ for some integer $n \geq 0$.  Note that this exists for any choice of  $\frac{2}{3} > \varepsilon > 0$:

We want to display an $n \in \mathbb{Z}_{\geq 0}$ such that:  

$$n\varepsilon \leq \frac{2}{3} - 2d < (n+1)\varepsilon$$

We have set $d = \frac{\varepsilon}{8}$, so the inequality $n\varepsilon \leq \frac{2}{3} - 2d < (n+1)\varepsilon$ is equivalent to: $$n\varepsilon \leq \frac{2}{3} - \frac{\varepsilon}{4} < (n+1)\varepsilon$$

Set $n = \left \lfloor{\frac{2}{3\varepsilon} - \frac{1}{4}}\right \rfloor $.  Since $\varepsilon > 0$ then this $n$ will always be finite.

\textbf{Case 2.I.1} $0 \leq \frac{2}{3} - 2d < \varepsilon$ 

Then, voters in profile $u_1$ with preferences $\begin{pmatrix}  x \\ y \\ z \end{pmatrix}$ will misreport their preferences as $\begin{pmatrix}  x \\ z \\ y \end{pmatrix}$, and voters in profile $u_1$ with preferences $\begin{pmatrix}  z \\ x \\ y \end{pmatrix}$ will also misreport their preferences as $\begin{pmatrix}  x \\ z \\ y \end{pmatrix}$, inducing $u_2$ and evaluation $(E2)$, and improving their outcome from $y$ to $x$.  This is a coalition size of $2(\frac{1}{3} - d) < \varepsilon$.  This would contradict WSP, meaning that the profile of $u_2$ could not evaluate to $x$.

\smallskip

\textbf{Case 2.I.2} $\varepsilon \leq \frac{2}{3} - 2d < 2\varepsilon$ 

Now, for convenience, denote $\frac{1}{3} - d$ as $q$.  Recall that:

$$F(u_1) = y$$

And, we are supposing that:

$$ F(u_2) = x$$

Consider the following profile:

$$u' = \begin{tabular}{ c c c }
q &  (1/3) + 2d  & q \\
\hline
 x  & y  & x\\ 
y  & z & z\\  
z  & x &y   
\end{tabular} $$

What can $F(u')$ evaluate to?  Suppose that $F(u') \neq y$.  Then, in profile $u_1$, those with preferences $\begin{pmatrix}  z \\ x \\ y \end{pmatrix}$ would misreport as $\begin{pmatrix}  x \\ z \\ y \end{pmatrix}$ with coalition size $q < \varepsilon$, which would improve their outcome from $y$.  This would contradict WSP, so we conclude that $F(u') = y$.

But then (as an assumption for \textbf{2.I}), we are assuming that $F(u_2)$ evaluates to $x$.  Given this, in $u'$, people with preferences $\begin{pmatrix}  x \\ y \\ z \end{pmatrix}$ would misreport their preferences as $\begin{pmatrix}  x \\ z \\ y \end{pmatrix}$ (for a coalition of size $q$), inducing improved outcome $x$, and contradicting WSP.  In this case, it cannot be true that $F(u_2)$ evaluates to $x$.

\smallskip

\textbf{Case 2.I.n+1} $n\varepsilon \leq 2q < (n + 1)\varepsilon$ (for $n \geq 2)$

Again, recall that:

$$F(u_1) = y$$

We are supposing that:

$$ F(u_2) = x$$

Let $r = \left \lceil{\frac{n+1}{2}}\right \rceil$.  Note that $r \geq \frac{n+1}{2}$.

Consider the set of profiles:
$$u^{(j)} = \begin{tabular}{ c c c c}
q &  (1/3) + 2d  & (r - j)(q/r) & j(q/r)\\
\hline
 x  & y  & z  & x\\ 
y  & z & x & z\\  
z  & x &y  & y 
\end{tabular}$$

(where integer $j$ iterates from $0$ to $r - 1$)

Note that when $j = 0$, we have:

$$u^{(0)} = \begin{tabular}{ c c c }
q &  (1/3) + 2d  & q\\
\hline
 x  & y  & z \\ 
y  & z & x\\  
z  & x &y 
\end{tabular}$$

This profile is exactly equivalent to $u_1$, meaning that $F(u^{(0)}) = y$.

Then, consider 

$$u^{(1)} = \begin{tabular}{ c c c c}
q &  (1/3) + 2d  & (r - 1)(q/r) & (q/r)\\
\hline
 x  & y  & z  & x\\ 
y  & z & x & z\\  
z  & x &y  & y 
\end{tabular}$$

If $F(u^{(1)}) \neq y$, then in profile $u^{(0)}$, those with preferences $\begin{pmatrix} z \\ x \\ y \end{pmatrix}$ would form a coalition of size $\frac{q}{r}$ (as $\frac{q}{r} \leq  \frac{2(\frac{1}{3} - d)}{n + 1} < \varepsilon$) and misreport as $\begin{pmatrix} x \\ z \\ y \end{pmatrix}$, creating profile $u^{(1)}$, and inducing an outcome that is better than $y$ for them.  This would contradict $WSP$, so $F(u^{(1)}) = y$.

Now, assume that for some general $0 < j < r - 1$, $F(u^{(j)}) = y$.  We want to show that $F(u^{(j + 1)}) = y$.

So, we have:

$$u^{(j)} = \begin{tabular}{ c c c c}
q &  (1/3) + 2d  & (r - j)(q/r) & j(q/r)\\
\hline
 x  & y  & z  & x\\ 
y  & z & x & z\\  
z  & x &y  & y 
\end{tabular}$$

$$u^{(j+ 1)} = \begin{tabular}{ c c c c}
q &  (1/3) + 2d  & (r - j - 1)(q/r) & (j + 1)(q/r)\\
\hline
 x  & y  & z  & x\\ 
y  & z & x & z\\  
z  & x &y  & y 
\end{tabular}$$

Suppose that $F(u^{(j + 1)}) \neq y$.  But then, voters in $u^{(j)}$ with preferences 
$\begin{pmatrix} z \\ x \\ y \end{pmatrix}$ would form a coalition of size $\frac{q}{r}$ and misreport as $\begin{pmatrix} x \\ z \\ y \end{pmatrix}$, creating profile $u^{(j +1)}$, and improving their outcome.  This means that $F(u^{(j+ 1)}) = y$.

A similar logic applies for all $j$, so we have that $F(u^{(j)}) = y$.  In particular, we have that:

$$F(u^{(r - 1)}) = F \left( \begin{tabular}{ c c c c}
q &  (1/3) + 2d  & (q/r) & (r - 1)(q/r)\\
\hline
 x  & y  & z  & x\\ 
y  & z & x & z\\  
z  & x &y  & y 
\end{tabular} \right) = y \quad (E3)$$

From the other direction, consider the set of profiles:

$$u_{(k)} = \begin{tabular}{ c c c }
k(q/r) &  (1/3) + 2d  & (2r - k)(q/r) \\
\hline
 x  & y  &  x\\ 
y  & z & z\\  
z  & x &y 
\end{tabular}$$

(where integer $k$ iterates from $0$ to $r$).

Note that when $k = 0$, we have the following profile:

$$u_{(0)} = \begin{tabular}{ c c }
 (1/3) + 2d  & 2q \\
\hline
 y  &  x\\ 
 z & z\\  
 x &y 
\end{tabular}$$

By the property A, this is equivalent to the profile $u_2$.  So, we have that $F(u_{(0)}) = x$.  

Consider the next profile:

$$u_{(1)} = \begin{tabular}{ c c c }
(q/r) &  (1/3) + 2d  & (2r - 1)(q/r) \\
\hline
 x  & y  &  x\\ 
y  & z & z\\  
z  & x &y 
\end{tabular}$$

Suppose that $F(u_{(1)}) \neq x$.  But then, those in profile $u_{(1)}$ with preferences $\begin{pmatrix} x \\ y \\ z\end{pmatrix}$ would misreport as $\begin{pmatrix} x \\ z \\ y \end{pmatrix}$ (coalition size $\frac{q}{r}$), inducing profile $u_{(0)}$ and improved outcome $x$.  So, $F(u_{(1)}) = x$.

Again, this logic will apply for all $k$.  Assume for some general $k < r$ that $F(u_{(k)}) = x$.  So then, we want to show that $F(u_{(k+1)}) = x$.

$$u_{(k)} = \begin{tabular}{ c c c }
k(q/r) &  (1/3) + 2d  & (2r - k)(q/r) \\
\hline
 x  & y  &  x\\ 
y  & z & z\\  
z  & x &y 
\end{tabular}$$

$$u_{(k + 1)} = \begin{tabular}{ c c c }
(k + 1)(q/r) &  (1/3) + 2d  & (2r - k - 1)(q/r) \\
\hline
 x  & y  &  x\\ 
y  & z & z\\  
z  & x &y 
\end{tabular}$$

Suppose that $F(u_{(k + 1)}) \neq x$.  But then, those in profile $u_{(k+1)}$ with preferences $\begin{pmatrix} x \\ y \\ z\end{pmatrix}$ would misreport as $\begin{pmatrix} x \\ z \\ y \end{pmatrix}$ (coalition size $\frac{q}{r}$), inducing profile $u_{(k)}$ and improved outcome $x$.  So, $F(u_{(k + 1)}) = x$.  

This means that in particular, we have:

$$ F(u_{(r)}) = F \left( \begin{tabular}{ c c c }
q &  (1/3) + 2d  & q \\
\hline
 x  & y  &  x\\ 
y  & z & z\\  
z  & x &y 
\end{tabular} \right) = x$$

Note that for the evaluation $F(u^{(r - 1)})$ in $(E3)$ above, those with preferences $\begin{pmatrix} z \\ x \\ y\end{pmatrix}$ would misreport their preferences as $\begin{pmatrix} x \\ z \\ y\end{pmatrix}$ (with coalition size $\frac{q}{r}$).  This would induce profile $u_{(r)}$ and outcome $x$, contradicting WSP.  So, this case also fails.

This means that for any $\varepsilon$, for $F$ to satisfy the properties P, A, N, and WSP, it is impossible for $F$ to evaluate as $x$ on profile $u_2$.

\textbf{2.II}  Suppose that the profile $u_2$ has the following evaluation.  
$$F \left(\begin{tabular}{ c c c }
(1/3) - d &  (1/3) + 2d  & (1/3) - d\\
\hline
 x  & y  & x\\ 
z  & z & z\\  
y  & x &y   
\end{tabular} \right) = y  $$

By the property $A$, the profile $u_2$ is equivalent to the following:

$$ u_2 ' = \begin{tabular}{ c c  }
(2/3) - 2d &  (1/3) + 2d  \\
\hline
 x  & y  \\ 
z  & z \\  
y  & x   
\end{tabular} $$

This means that the following evaluation is also true:

$$F (u_2') = y \quad (E4)$$

Then, we apply the permutation $\sigma^{*}$, where $\sigma^{*}(x) = y, \sigma^{*}(y) = x, \sigma^{*}(z) = z$. By $A$ and $N$, we create the following profile and evaluation:

$$ u_3 = \begin{tabular}{ c c  }
(2/3) - 2d &  (1/3) + 2d  \\
\hline
 y  & x  \\ 
z  & z \\  
x  & y   
\end{tabular}$$

$$F(u_3) = x \quad (E5)$$

Recall that in the setup for this step, we assumed that $d < \frac{1}{12}$.  This is still the case, as we have set $d = \frac{\varepsilon}{8}$. Therefore, $(2/3) - 2d > \frac{1}{2} > (1/3) + 2d$.

Additionally, there exists $m \in \mathbb{Z}_{\geq 0}$ such that $m\varepsilon \leq \frac{1}{3} - 4d < (m+1)\varepsilon$.  Because $\varepsilon < \frac{2}{3}$, we can set $m = \left \lfloor{\frac{1}{3\varepsilon} - \frac{1}{2}}\right \rfloor$.  Since $0<\varepsilon < \frac{2}{3}$, then $m$ will always be greater than or equal to zero and finite.

For convenience, in the following steps, denote $\frac{2}{3} - 2d$ as $D$, and $\frac{1}{3} + 2d$ as $1 - D$.  Also, note that $\frac{1}{2} < D < \frac{2}{3}$, and $2D - 1 = \frac{1}{3} - 4d$.

\smallskip

\textbf{Case 2.II.1}:  $0 \leq \frac{1}{3} - 4d < \varepsilon$ 

In profile $u_3$, voters with preferences $\begin{pmatrix} y \\ z \\ x \end{pmatrix}$ would form a coalition of size $(2/3) - 2d - ((1/3) + 2d) = \frac{1}{3} - 4d = 2D - 1$, and misrepresent their preferences as $\begin{pmatrix} x \\ z \\ y \end{pmatrix}$, inducing $(E4)$, and getting them improved outcome $y$.  This would violate WSP. For this case, to satisfy the properties the we want, it cannot be true that $F(u_2) = y$.

\smallskip

\textbf{Case 2.II.2}:  $\varepsilon \leq \frac{1}{3} - 4d < 2\varepsilon$ (i.e., $\varepsilon \leq 2D - 1 < 2\varepsilon$)

Again, recall that we are assuming:  $$ F(u_2') = y \quad (E4)$$

And, from $(E4)$ and the properties A, N, it follows that:

$$ F(u_3) = x \quad (E5)$$

Consider the following profile:

$$u_4 = \begin{tabular}{ c c } D - k & (1 - D) + k \\ \hline x & y \\z & z \\ y & x \end{tabular} $$

(where $k = \frac{2D-1}{2}$, so $k < \varepsilon$.  Also, note that in this case, $D - k = \frac{1}{2}$, and $(1 - D) + k = \frac{1}{2}$).

We want to consider what $F(u_4)$ can evaluate to.  Suppose that $F(u_4) \neq y$.  But then, in profile $u_2'$, voters with preferences $\begin{pmatrix} x \\ z \\ y \end{pmatrix}$ would form a coalition of size $k < \varepsilon$, and misreport as $\begin{pmatrix} y \\ z \\ x \end{pmatrix}$, inducing $u_4$ and an improved outcome over $y$.  This would contradict WSP, so, $F(u_4) = y$.

But then, voters in profile $u_3$ with preferences $\begin{pmatrix} y \\ z \\ x \end{pmatrix}$ would form a coalition of size $k < \varepsilon$, and misreport as $\begin{pmatrix} x \\ z \\ y \end{pmatrix}$, inducing profile $u_4$ and improved outcome $y$.  This would contradict WSP.  So, again, for this case, to satisfy the properties the we want, it cannot be true that $F(u_2) = y$.

\textbf{Case 2.II.m+1}
$m\varepsilon \leq \frac{1}{3} - 4d < (m+1)\varepsilon$ (for $m \geq 2)$

Again, recall that we are assuming:  $$ F(u_2') = y \quad (E4)$$

And, from $(E4)$ and the properties A, N, it follows that:

$$ F(u_3) = x \quad (E5)$$

Now, consider the following profiles:

$$u^{(j)} = \begin{tabular}{ c c } D - j$k'$ & (1 - D) + j$k'$ \\ \hline x & y \\z & z \\ y & x \end{tabular} $$

(where $j$ iterates from $1$ to $m$, and $k' = \frac{2D - 1}{m + 1}$.  And, note that this definition of $u^{(j)}$ is local to \textbf{Case 2.II.m+1}).

We first consider the evaluation of $u^{(1)}$.  Suppose that $F(u^{(1)}) \neq y$.  But then, in profile $u_2'$, voters with preferences $\begin{pmatrix} x \\ z \\ y \end{pmatrix}$ would form a coalition of size $k'$, and misreport as $\begin{pmatrix} y \\ z \\ x \end{pmatrix}$, inducing profile $u^{(1)}$ and an improved outcome over $y$.  This would contradict WSP, so, $F(u^{(1)}) = y$.

For a general $j$, assume that $F(u^{(j)}) = y$.  By a similar logic to before, we can show that $F(u^{(j+1)}) = y$.  Suppose that $F(u^{(j+1)}) \neq y$.  Then, in profile $u^{(j)}$, voters with preferences $\begin{pmatrix} x \\ z \\ y \end{pmatrix}$ would form a coalition of size $k'$, and misreport as $\begin{pmatrix} y \\ z \\ x \end{pmatrix}$, inducing profile $u^{(j+1)}$ and an improved outcome over $y$.  This would contradict WSP, so, $F(u^{(j+1)}) = y$, for any $j$.

We have the following profile and evaluation:

$$u^{(m)} = \begin{tabular}{ c c } D - m$k'$ & (1 - D) + m$k'$ \\ \hline x & y \\z & z \\ y & x \end{tabular} $$

$$F(u^{(m)}) = y$$

But then, note that in profile $u_3$, voters who have preferences $\begin{pmatrix} y \\ z \\ x \end{pmatrix}$ would form a coalition of size $k'$, and misreport as $\begin{pmatrix} x \\ z \\ y \end{pmatrix}$, inducing profile $u^{(m)}$ and improved outcome $y$.  

Clearly, $D - (1 - D + mk') = 2D - 1 - mk' = \frac{(m + 1)(2D - 1) - m(2D - 1)}{m + 1} = k' < \varepsilon$. 

This would contradict WSP.  So, for a general $\frac{1}{3} - 4d$, then $F$ will not satisfy the properties.  Thus, it is impossible for $F$ to evaluate as $y$ on $u_2$.

\textbf{2.III}  Finally, suppose that the profile $u_2$ has the following evaluation.  
$$F \left(\begin{tabular}{ c c c }
(1/3) - d &  (1/3) + 2d  & (1/3) - d\\
\hline
 x  & y  & x\\ 
z  & z & z\\  
y  & x &y   
\end{tabular} \right) = z \quad (E6) $$

Equivalently, by the property A,

$$u_2' = \begin{tabular}{ c c }
(2/3) - 2d &  (1/3) + 2d\\
\hline
 x  & y \\ 
z  & z \\  
y  & x   
\end{tabular}$$
$$F (u_2') = z $$

Consider the profile:

$$ u_5 = \begin{tabular}{ c c }
(2/3) - 2d &  (1/3) + 2d\\
\hline
 x  & z \\ 
z  & x \\  
y  & y   
\end{tabular}$$

We can show that it must be true that:

$$F(u_5) = z \quad (E7)$$

\textbf{2.III.0} (set up) - showing that $F(u_5) = z$

We will use an iterative approach to show that $F(u_5)$ must equal $z$ for any $d$ and $\varepsilon$.  The cases will be based on the following inequality:  $$h\varepsilon \leq \frac{1}{3} + 2d < (h +1)\varepsilon$$  Note that there exists some integer $h \geq 0$ where the weight $\frac{1}{3} + 2d$ fulfills the inequality.  We have set $d = \frac{\varepsilon}{8}$, so we can let $h = \left \lfloor{\frac{1}{3\varepsilon} + \frac{1}{4}}\right \rfloor$.

\textbf{Case 2.III.0.1}: $0 \leq \frac{1}{3} + 2d < \varepsilon$

So, suppose that $F(u_5) \neq z$.  Then, in profile $u_5$, voters with preferences $\begin{pmatrix} z \\ x \\ y \end{pmatrix}$ would form a coalition of size $\frac{1}{3} + 2d$ and misreport as $\begin{pmatrix} y \\ z \\ x \end{pmatrix}$, inducing profile $u_2'$ and improved outcome $z$.  This would contradict WSP, so $F(u_5) = z$.

\textbf{Case 2.III.0.2}: $\varepsilon \leq \frac{1}{3} + 2d < 2\varepsilon$

Again, we proceed by contradiction, and suppose that $F(u_5) \neq z$.  Consider the following profile:

$$ u_{5}^* = \begin{tabular}{ c c c}
(2/3) - 2d & c & (1/3) + 2d - c \\
\hline
 x  & z & y\\ 
z  & x & z\\  
y  & y  &x 
\end{tabular}$$

(Here, we let $c = \frac{\frac{1}{3} + 2d}{2}$)

So, what can $F(u_{5}^*)$ evaluate to?  Suppose that $F(u_{5}^*) \neq z$.  But then, voters in $u_{5}^*$ with preferences $\begin{pmatrix} z \\ x \\ y \end{pmatrix}$ would form a coalition of size $c < \varepsilon$, and misreport their preferences as $\begin{pmatrix} y \\ z \\ x \end{pmatrix}$, inducing profile $u_2'$ and improved outcome $z$.  This would contradict WSP, so $F(u_{5}^*) = z$.

Suppose that $F(u_{5}) \neq z$.  But then, voters in $u_{5}$ with preferences $\begin{pmatrix} z \\ x \\ y \end{pmatrix}$ would form a coalition of size $c < \varepsilon$, and misreport their preferences as $\begin{pmatrix} y \\ z \\ x \end{pmatrix}$, inducing profile $u_{5}^*$ and improved outcome $z$.  This would contradict WSP, so $F(u_{5}) = z$.

\textbf{Case 2.III.0.h+1}: $h\varepsilon \leq \frac{1}{3} + 2d < (h + 1)\varepsilon$ (for $h \geq 2)$

Consider the following profiles:

$$ u^{(j)}_5 = \begin{tabular}{ c c c}
(2/3) - 2d & j$c'$ & (1/3) + 2d - j$c'$ \\
\hline
 x  & z & y\\ 
z  & x & z\\  
y  & y  &x 
\end{tabular}$$
(Here, $c' = \frac{\frac{1}{3} + 2d}{h+1}$, and $j$ iterates from $0$ to $h$.)

When $j = 0$, then this profile is just equivalent to $u_2'$.  So, we know that $F(u^{(0)}_5) = z$.  Then, we want to know what the evaluation $F(u^{(1)}_5)$ is.  Suppose that $F(u^{(1)}_5) \neq z$.  But then, voters in $u^{(1)}_5$ with preferences $\begin{pmatrix} z \\ x \\ y \end{pmatrix}$ would form a coalition of size $c' < \varepsilon$, and misreport their preferences as $\begin{pmatrix} y \\ z \\ x \end{pmatrix}$, inducing profile $u^{(0)}_5$ and improved outcome $z$.  This would contradict WSP, so $F(u^{(1)}_5) = z$.  A similar logic applies for all $j$, so this means that $F(u^{(h)}_5) = z$.  Assume that $F(u^{(j)}_5) = z$ for some general $j$.  Then, we want to know what the evaluation $F(u^{(j+1)}_5)$ is.  Suppose that $F(u^{(j+1)}_5) \neq z$.  But then, voters in $u^{(j+1)}_5$ with preferences $\begin{pmatrix} z \\ x \\ y \end{pmatrix}$ would form a coalition of size $c' < \varepsilon$, and misreport their preferences as $\begin{pmatrix} y \\ z \\ x \end{pmatrix}$, inducing profile $u^{(j)}_5$ and improved outcome $z$.  This would contradict WSP, so $F(u^{(j+1)}_5) = z$.

But then this means that $F(u_5)$ must equal $z$:  otherwise, if $F(u_5) \neq z$, then voters in profile $u_5$ with preferences $\begin{pmatrix} z \\ x \\ y \end{pmatrix}$ would deviate with a coalition size of $c'$, and misreport their preferences as $\begin{pmatrix} y \\ z \\ x \end{pmatrix}$, inducing profile $u^{(h)}_5$ and outcome $z$.  Therefore, $F(u_5)= z$.

We have that $F(u_5) = z$, and we see that is true for a general $\varepsilon$.  This concludes \textbf{Step 2.III.0}.  

\smallskip

Apply the following permutation: $\sigma^{**}(x) = z, \sigma^{**}(y) = y$, $\sigma^{**}(z) = x$ to $u_5$.  With N and A, this gives us the following profile and evaluation:

$$ u_6 = \begin{tabular}{ c c }
(2/3) - 2d &  (1/3) + 2d\\
\hline
 z  & x \\ 
x  & z \\  
y  & y   
\end{tabular}$$

$$F(u_6) = x \quad (E8)$$

As in \textbf{2.II}, denote $D = \frac{2}{3} - 2d$, and set $m$ as it was defined before.

\textbf{Case 2.III.1} $0 \leq 2D - 1 < \varepsilon$

In evaluation $(E7)$ of profile $u_5$, voters with preferences $\begin{pmatrix} x \\ z \\ y \end{pmatrix}$ would form a coalition of size $2D - 1$, and misrepresent their preferences as $\begin{pmatrix} z \\ x \\ y \end{pmatrix}$, inducing profile $u_6$ and evaluation $(E8)$, and achieving improved outcome $x$.  This would violate WSP, meaning that for this case, $F(u_2) \neq z$.

\textbf{Case 2.III.2} $\varepsilon \leq 2D - 1 < 2\varepsilon$

Again, recall that we are assuming:  $$ F(u_2') = z $$

We also have:

$$ F(u_5) = z \quad (E7) $$

$$ F(u_6) = x  \quad (E8) $$

Consider the following profile:

$$u_7 = \begin{tabular}{ c c } D - k & (1 - D) + k \\ \hline x & z \\z & x \\ y & y \end{tabular} $$

(where $k = \frac{2D-1}{2}$, so $k < \varepsilon$.  Again, note that in this case, $D - k = \frac{1}{2}$, and $(1 - D) + k = \frac{1}{2}$).

We want to consider the evaluation $F(u_7)$.  Suppose that $F(u_7) = x$ (also note that $y$ is Pareto dominated on this profile, and cannot be the result).  But then, in profile $u_5$, voters with preferences $\begin{pmatrix} x \\ z \\ y \end{pmatrix}$ would form a coalition of size $k$, and misreport as $\begin{pmatrix} z \\ x \\ y \end{pmatrix}$, inducing $u_7$ and an improved outcome over $z$.  This would contradict WSP, so, $F(u_7) = z$.

But then, voters in profile $u_6$ with preferences $\begin{pmatrix} z \\ x \\ y \end{pmatrix}$ would form a coalition of size $k$, and misreport as $\begin{pmatrix} x \\ z \\ y \end{pmatrix}$, inducing profile $u_7$ and improved outcome $z$.  This would contradict WSP.  This means that $F(u_7)$ cannot equal $z$.  But, this violates the initial hypothesis that $F(u_2) = z$.  So, again, for this case, to satisfy the properties that we want, it cannot be true that $F(u_2) = z$.

\textbf{Case 2.III.m+1} $m\varepsilon \leq 2D - 1 < (m+1)\varepsilon$ (for $m \geq 2)$

Again, recall that we are assuming:  $$ F(u_2') = z $$

And, we have that:

$$ F(u_5) = z \quad (E7) $$

$$ F(u_6) = x \quad (E8)$$

Now, consider the following profiles:

$$u^{(j)} = \begin{tabular}{ c c } D - j$k'$ & (1 - D) + j$k'$ \\ \hline x & z \\z & x \\ y & y \end{tabular} $$

(where $j$ iterates from $1$ to $m$, and $k' = \frac{2D - 1}{m + 1}$.  Also note that this definition of $u^{(j)}$ is local to \textbf{Case 2.III.m+1}.)

First consider the evaluation when we have $u^{(1)}$.  Suppose that $F(u^{(1)}) = x$.  But then, in $(E7)$, voters in profile $u_5$ with preferences $\begin{pmatrix} x \\ z \\ y \end{pmatrix}$ would form a coalition of size $k'$, and misreport as $\begin{pmatrix} z \\ x \\ y \end{pmatrix}$, inducing profile $u^{(1)}$ and an improved outcome $x$.  This would contradict WSP, so, $F(u^{(1)}) = z$.

For a general $j$, assume that $F(u^{(j)}) = z$.  By a similar logic to before, we can show that $F(u^{(j+1)}) = z$.  Suppose that $F(u^{(j+1)}) = x$.  So then in profile $u^{(j)}$, voters with preferences $\begin{pmatrix} x \\ z \\ y \end{pmatrix}$ would form a coalition of size $k'$, and misreport as $\begin{pmatrix} z \\ x \\ y \end{pmatrix}$, inducing profile $u^{(j+1)}$ and an improved outcome $x$ over $z$.  This would contradict WSP, so, $F(u^{(j+1)}) = z$, for any $j$.

This gives us the following profile and evaluation:

$$u^{(m)} = \begin{tabular}{ c c } D - m$k'$ & (1 - D) + m$k'$ \\ \hline x & z \\z & x \\ y & y \end{tabular} $$

$$F(u^{(m)}) = z$$

But then, note that in profile $u_6$, the voters with preferences $\begin{pmatrix} z \\ x \\ y \end{pmatrix}$ would form a coalition of size $k'$, and misreport as $\begin{pmatrix} x \\ z \\ y \end{pmatrix}$, inducing profile $u^{(m)}$ and improved outcome $z$.  Again, the coalition size is less than $\varepsilon$, as in \textbf{2.II.m+1}.

This would contradict WSP.  This in turn contradicts the original hypothesis that $F(u_2) = z$.  

This concludes \textbf{Step 2}.  We see that for a general $\varepsilon$, knowing that $F(u_1) = y$, then there is nothing that $F$ can evaluate to on $u_2$.  This means that there is no voting rule $F$ that satisfies the properties P, A, N, and WSP on an expansion of the Condorcet domain.

\newpage

\section*{Step 3}
We first consider the following rich domain:

$$U^{I} = \Bigg\{ \begin{tabular}{ c c c c}
 x  & y & y & z\\ 
y  & z & x & y\\  
z  & x & z & x    
\end{tabular} \Bigg\} $$

Consider the voting rule $F$ satisfying P, A, N, and WSP on $U^{I}$.  We want to show that there is no voting rule $F \neq F^C$ that satisfies these axioms.  Fix $\varepsilon >0$. 

Consider the generic profile:

$$u_{.}^{I} =  \begin{tabular}{ c c c c}
a & b & c & 1 - a - b - c\\
\hline
 x  & y & y & z\\ 
y  & z & x & y\\  
z  & x & z & x    
\end{tabular} $$

(Note that the profiles and weights here have no relation to the weights and profiles that were named in previous steps).

\textbf{3.I.1}  Suppose that the result of Condorcet $F^{C}(u_{.}^{I})$ is $x$. So, $a > \frac{1}{2}$.

\textbf{3.I.1.1}
Suppose that:

$$F(u_{.}^{I}) = y \quad (I1)$$  

Consider the following profile:

$$u^{I}_1 = \begin{tabular}{ c c }
a &  1 - a\\
\hline
 x  & y \\ 
y  & x \\  
z  & z    
\end{tabular}$$

\textbf{3.I.1.1.0} We want to show that:

$$F(u^{I}_1) = y $$

Assume that $a$ is fixed, and the conditions for the following cases are based on $c$ (and implicitly on $b$).

\textbf{Case 3.I.1.1.0.1:} $0 \leq 1 - a - c < \varepsilon$

If $F(u^{I}_1)$ was not $y$, then those in profile $u^{I}_1$ with preferences $\begin{pmatrix} y \\ x \\ z \end{pmatrix}$ would form a coalition of size $1 - a - c$, where a weight of $b$ would misrepresent as $\begin{pmatrix} y \\ z \\ x \end{pmatrix}$, and a weight of $1 - a - b - c$ would misrepresent as $\begin{pmatrix} z \\ y \\ x \end{pmatrix}$, inducing profile $u_{.}^{I}$ and evaluation $(I1)$, leading to improved result $y$.  This would contradict WSP.  So, when $1 - a - c < \varepsilon$, $F(u^{I}_1) = y$.  

\smallskip

\textbf{Case 3.I.1.1.0.2:} $\varepsilon \leq 1 - a - c < 2\varepsilon$

Again, by assumption for this step, we are supposing that: 

$$ F (u_.^I)= y \quad (I1)$$

Consider the following profile:

$$u^{I}_2 = \begin{tabular}{ c c c c}
a & k & 1 - a - k - m & m\\
\hline
 x  & y & y & z\\ 
y  & z & x & y\\  
z  & x & z & x    
\end{tabular}$$

(Here, $k = \frac{b}{2}; m = \frac{1 - a - b - c}{2}$).

\smallskip

Suppose that $F(u^{I}_2) \neq y$. 

Then, people with preferences $\begin{pmatrix} y \\ x \\ z \end{pmatrix}$ in $u^{I}_2$ would want to induce initial profile $u_{.}^{I}$.  So, a group of size $b - k$ would misrepresent their preferences as $\begin{pmatrix} y \\ z \\ x \end{pmatrix}$, and a group of size $1 - a - b - c - m$ would misrepresent their preferences as $\begin{pmatrix} z \\y \\ x \end{pmatrix}$.  This would create profile $u_{.}^{I}$, and improved outcome $y$.  Note that the coalition size is less than $\varepsilon$:

$$b - k + 1 - a - b - c - m = \frac{1 - a - c}{2} < \varepsilon$$

So, to satisfy WSP, this tells us that $F(u^{I}_2) = y$.

Consider renaming the weights on $u^{I}_2$ in the following manner.  

Where:

$$ a^{*} = a; \quad b^{*} = k; \quad c^{*} = 1 - a - k - m; \quad 1 - a^{*} - b^{*} - c^{*} = m.$$

$$u^{I}_2 = \begin{tabular}{ c c c c}
$a^{*}$ & $b^{*}$ & $c^{*}$ & $1 - a^{*} - b^{*} - c^{*}$\\
\hline
 x  & y & y & z\\ 
y  & z & x & y\\  
z  & x & z & x    
\end{tabular}$$

This is still the same profile but with different labels on the weights.  We see that $b^{*} + 1 - a^{*} - b^{*} - c^{*} = k + m < \varepsilon$.  So, knowing the $F(u^{I}_2) = y$, we see that $u^{I}_2$ is just an instance of \textbf{Case 3.I.1.1.0.1}.  This tells us that $F(u^{I}_1) = y$.

Now, for the inductive hypothesis, we assume that all cases up to \textbf{Case 3.I.1.1.0.n} hold.  In other words, given some profile:

$$ u^{*} = \begin{tabular}{ c c c c }
$a$ & $b^{*}$ & $c^*$ & $1 - a - b^{*} - c^*$\\
\hline
 x  & y & y & z\\ 
y  & z & x & y\\  
z  & x & z & x    
\end{tabular} $$

where the following conditions are true:

$$ F(u^{*}) = y$$

$$(n- 1)\varepsilon \leq  1 - a - c^*  < n\varepsilon \quad (W1)$$

Then this implies that $F(u^{I}_1) = y$.

Now, we consider the following case:

\textbf{Case 3.I.1.1.0.n+1:} $n\varepsilon \leq 1 - a - c < (n+1)\varepsilon$

We are still supposing that:

$$F(u_.^{I}) = y \quad (I1)$$ 

Consider the following profile:

$$u^{I}_{(n)} = \begin{tabular}{ c c c c}
a & k$'$ & 1 - a - k$'$ - m$'$ & m$'$\\
\hline
 x  & y & y & z\\ 
y  & z & x & y\\  
z  & x & z & x    
\end{tabular} $$

(Here, $k' = \frac{nb}{n+1}$, $m' = \frac{n(1 - a - b - c)}{n+1}$).

We consider the result $F(u^{I}_{(n)})$. Suppose that $F(u^{I}_{(n)}) \neq y$.  But then, voters with preferences $\begin{pmatrix} y \\ x \\ z \end{pmatrix}$ in $u^{I}_{(n)}$ would want to induce initial profile $u_{.}^{I}$.  A weight of $b - k'$ would misreport as $\begin{pmatrix} y \\ z \\ x \end{pmatrix}$ and a weight of $1 - a - b - c - m'$ would misreport their preferences as $\begin{pmatrix} z\\ y \\ x \end{pmatrix}$.  This would create profile $u_{.}^{I}$ and improved result $y$, contradicting WSP.  So, this tells us that $F(u^{I}_{(n)}) = y$.

But then, note that the weights on profile $u^{I}_{(n)}$ fulfill the following:

$$ F(u^{I}_{(n)}) = y$$

$$ (n - 1)\varepsilon \leq k' + m' = \frac{n(1 - a - c)}{n + 1} < n\varepsilon$$

So, this we see that profile $u_{(n)}^I$ is a case of a profile $u^{*}$ that satisfies a case of \textbf{3.I.1.1.0.n}, where $b^* = k', c^{*} = 1 - a - k' - m', 1 - a - b^* - c^* = m'$.  By the inductive hypothesis, this tells us that:

$$ F(u^{I}_1) = y$$ 

This is true in general.  This tells us that for any $a > \frac{1}{2}$, and for any $c$ that satisfies $1 - a - c > 0$, if the winner of $F$ on $u^{I}_.$ is $y$, then it also follows that: 

$$F\left( \begin{tabular}{ c c }
a &  1 - a\\
\hline
 x  & y \\ 
y  & x \\  
z  & z    \end{tabular} \right) = y \quad (I2)$$

\textbf{3.I.1.1.1} Now, we will show that a voting rule $F \neq F^C$ that satisfies the axioms, where $F(u^{I}_1) = y$, is impossible.

Apply permutation $\sigma^*$ to profile $u^{I}_1$, where $\sigma^*(x) = y, \sigma^*(y) = x$, and $\sigma^*(z) = z$.

This gives us the following profile and evaluation:

$$  u^{I}_3 =  \begin{tabular}{ c c }
a &  1 - a\\
\hline
 y  & x \\ 
x  & y \\  
z  & z 
\end{tabular}$$

$$ F(u^{I}_3) = x \quad (I3)$$

\textbf{Case 3.I.1.1.1.1}  $0 \leq |2a - 1| < \varepsilon$

If $a - (1 - a) < \varepsilon$, then voters in profile $u^{I}_3$ with preferences $\begin{pmatrix} y \\ x \\ z \end{pmatrix}$ would form a coalition of size $2a - 1$, and misreport their preferences as $\begin{pmatrix} x\\ y \\ z \end{pmatrix}$.  This would create profile $u^{I}_1$ and evaluation $(I2)$, giving improved result $y$.  

This would contradict WSP, meaning that on this range of $a$ and $b$, our original hypothesis that $F \neq F^{C}$ cannot be true.  So, this tells us that when $a^{*}$ fulfills the following inequality, then $F$ must agree with $F^C$:

$$ \frac{1}{2} < a^* < \frac{1}{2} + \frac{\varepsilon}{2} \quad (W2)$$

Then:

$$
F \left(
\begin{tabular}{ c c c }
$a^*$ & $1 - a^*$\\
\hline
 x & y \\ 
y & x \\  
z & z    
\end{tabular} \right) = x
$$

\textbf{Case 3.I.1.1.1.2}  $\varepsilon \leq |2a - 1| < 2\varepsilon$ (i.e., $\frac{\varepsilon}{2} + \frac{1}{2} < a \leq \varepsilon + \frac{1}{2}$)

Note that $a = \frac{1}{2} + \varepsilon - \delta$ for some $0 < \delta \leq \frac{\varepsilon}{2}$.

Again, we have:

$$ F(u^{I}_1) = y \quad (I2)$$

Now, consider the following profile:

$$u^{I}_4 = \begin{tabular}{ c c }
a - k &  1 - a + k\\
\hline
 x  & y \\ 
y  & x \\  
z  & z 
\end{tabular}$$

Where:  $$k = \frac{a - \frac{1}{2}}{2} = \frac{\varepsilon - \delta}{2} < \varepsilon$$

Note that $a - k = \frac{1}{2} + \frac{\varepsilon}{2} - \frac{\delta}{2} < \frac{1}{2} + \frac{\varepsilon}{2}$

Additionally, note that:

$$ \frac{\delta}{2} < \frac{\varepsilon}{4}$$

$$ \frac{\varepsilon}{4} - \frac{\delta}{2} > 0 $$

The weight on preferences $\begin{pmatrix} x \\ y \\ z \end{pmatrix}$ in profile $u^{I}_4$ also satisfies the following inequality:

$$a - k = \frac{1}{2} + \frac{\varepsilon}{2} - \frac{\delta}{2} = \frac{1}{2} + \frac{\varepsilon}{4} + \frac{\varepsilon}{4} - \frac{\delta}{2} > \frac{1}{2} + \frac{\varepsilon}{4} $$

This means that:

$$ \frac{1}{2} < a - k < \frac{1}{2} + \frac{\varepsilon}{2}$$

So, this weight is within the range where $F = F^{C}$ from \textbf{Case 3.I.1.1.1.1}.  So, $F(u^{I}_4) = x$.

But then, note that voters in profile $u^{I}_1$ with preferences $\begin{pmatrix} x\\ y \\ z \end{pmatrix}$ would misreport as $\begin{pmatrix} y\\ x \\ z \end{pmatrix}$ to induce profile $u^{I}_4$ and improved result $x$.  This again would contradict WSP, meaning that on this range of $(a, b)$, we have that $F = F^{C}$.

A similar logic will hold for all $a > \frac{1}{2}$.  So, in general, we see that our assumption for the evaluation $F(u^{I}_.)$ in \textbf{3.I.1.1} fails to hold, and we move on to \textbf{3.I.1.2}.

\textbf{3.I.1.2} Now we suppose that:

$$F(u_{.}^{I}) = z \quad (I4)$$

As before, we have:
$$u_{.}^{I}  =  \begin{tabular}{ c c c c}
a & b & c & 1 - a - b - c\\
\hline
 x  & y & y & z\\ 
y  & z & x & y\\  
z  & x & z & x    
\end{tabular} $$

Consider the following profile:

$$ u^I_4 = \begin{tabular}{ c c }
a &  1 - a\\
\hline
 x  & z \\ 
y  & y \\  
z  & x    
\end{tabular}$$

\textbf{3.I.1.2.0} Now we want to show that:

$$F(u^I_4) = z $$

Assume that $a$ is fixed, and the conditions for the following cases are based on $b$ and $c$.

\textbf{Case 3.I.1.2.0.1:} $0 \leq b + c < \varepsilon$

We consider the evaluation of $F(u^I_4)$.  By supposition, we have:

$$ F(u_.^I) = z \quad (I4)$$

Suppose that $F(u^I_4) \neq z$.  But then, voters in profile $u^I_4$ with preferences $\begin{pmatrix} z \\ y \\ x \end{pmatrix}$ would want to induce outcome $z$ and evaluation $(I4)$.  So, they would deviate, and misreport $\begin{pmatrix} y \\ z \\ x \end{pmatrix}$ with weight $b$, and misreport $\begin{pmatrix} y \\ x \\ z \end{pmatrix}$ with weight $c$, forming profile $u_.^I$, and result $z$.  So, this would improve their result, contradicting WSP.  For this case, we have that $F(u^I_4) = z$.

\textbf{Case 3.I.1.2.0.2:} $\varepsilon \leq b + c < 2\varepsilon$

We are still assuming that:

$$F(u_{.}^{I}) = z \quad (I4)$$
Consider the following profile:

$$ u^I_5 = \begin{tabular}{ c c c c}
a & k & m & 1 - a - k - m\\
\hline
 x  & y & y & z \\ 
y  & z & x & y \\  
z  & x & z & x    
\end{tabular} $$

Let $k = \frac{b}{2}$, and $m = \frac{c}{2}$.  Note that the definitions of $k$ and $m$ are local to this case.

Suppose that $F(u^I_5) \neq z$.  But then, voters in profile $(u^I_5)$ with preferences $\begin{pmatrix} z \\ y \\ x \end{pmatrix}$ would want to induce evaluation $(I4)$ and outcome $z$.  They would deviate, and misreport $\begin{pmatrix} y \\ z \\ x \end{pmatrix}$ with weight $b - k$, and misreport $\begin{pmatrix} y \\ x \\ z \end{pmatrix}$ with weight $c - m$.  This would create profile $u_{.}^{I}$ and evaluation $(I4)$, inducing outcome $z$.  So, this would improve their result, contradicting WSP.  This tells us that $F(u^I_5) = z$.

But then, we consider the following relabeling of the weights in profile $u^I_5$:  

$$ a = a; \quad b' = k; \quad c' = m; \quad 1 - a' - b' - c' = 1 - a - k - m$$

Note that $b' + c' < \varepsilon$.  We have shown that it must be true that $F(u^I_5) = z$.  So, profile $u^{I}_5$ is just a case of \textbf{3.I.1.2.0.1}, meaning that:

$$ F(u^I_4) = z$$

For the inductive hypothesis, we assume that all cases up to and including \textbf{Case 3.I.1.2.0.n} hold.  So, in other words, given some profile:

$$ u^{*} = \begin{tabular}{ c c c c}
$a$ & $b^*$ & $c^*$ & $1 - a - b^* - c^*$\\
\hline
 x  & y & y & z \\ 
y  & z & x & y \\  
z  & x & z & x    
\end{tabular} $$

where:

$$ F(u^{*}) = z$$

and the following inequality holds:

$$  (n-1)\varepsilon \leq b^* + c^* < n\varepsilon \quad (W3)$$

then, it is also true that $$ F(u^I_4) = z$$

\textbf{Case 3.I.1.2.0.n+1} $n\varepsilon \leq b + c < (n+1)\varepsilon$

Again we have:

$$ F(u_.^I) = z \quad (I4)$$

We want to show:

$$ F(u^I_4) = z$$

Consider the following profile:

$$ u^{I}_{(n)} = \begin{tabular}{ c c c c}
a & k & m & 1 - a - k - m\\
\hline
 x  & y & y & z \\ 
y  & z & x & y \\  
z  & x & z & x    
\end{tabular} $$

Here, we let $k = \frac{nb}{n + 1}$, and $m = \frac{nc}{n + 1}$.  Also note that this definition of $u^{I}_{(n)}$ is local to this case.

Suppose that $F(u^{I}_{(n)}) \neq z$. But then, voters with preferences $\begin{pmatrix} z \\ y \\ x \end{pmatrix}$ would want to induce profile $u_.^{I}$ and outcome $z$.

Voters in profile $u^{I}_{(n)}$ with preferences $\begin{pmatrix} z \\ y \\ x \end{pmatrix}$ would deviate, and misreport $\begin{pmatrix} y \\ z \\ x \end{pmatrix}$ with weight $b - k$, and misreport $\begin{pmatrix} y \\ x \\ z \end{pmatrix}$ with weight $c - m$.  This would be a coalition of size $\frac{b + c}{n + 1} < \varepsilon$.  So, this would create profile $u_.^I$ and evaluation $(I4)$, inducing outcome $z$.  This would result in an improved outcome for the $\begin{pmatrix} z \\ y \\ x \end{pmatrix}$ coalition, and would contradict WSP.  So, $F(u^{I}_{(n)}) = z$.

Then, note that the weights in profile $u^{I}_{(n)}$ fulfill inequality $(W3)$:

$$ (n - 1) \varepsilon \leq \frac{n(b + c)}{n + 1} < n \varepsilon$$

So, profile $u^{I}_{(n)}$ reduces to a profile $u^*$ that satisfies \textbf{Case 3.I.1.2.0.n} (where $b^* = k$ and $c^* = m$).  This implies that $F(u^I_4) = z$.  This is true in general.

$$ F(u^I_4) = z \quad (I5)$$

\textbf{3.I.1.2.1} Now, we will show that a voting rule $F \neq F^C$ that satisfies the axioms, where $F(u^{I}_4) = z$, is impossible.

Apply permutation $\sigma^{**}$ to $u^{I}_4$, where $\sigma^{**}(x) = z, \sigma^{**}(z) = x, \sigma^{**}(y) = y$.

$$ u^{I}_4 = \begin{tabular}{ c c }
a &  1 - a\\
\hline
 x  & z \\ 
y  & y \\  
z  & x    
\end{tabular}$$

$$ F(u^{I}_4) = z \quad (I6)$$

$$ u^{I}_5 = \begin{tabular}{ c c }
a &  1 - a\\
\hline
 z  & x \\ 
y  & y \\  
x  & z   
\end{tabular} = \begin{tabular}{ c c }
1 - a &  a\\
\hline
 x  & z \\ 
y  & y \\  
z  & x    
\end{tabular}$$

By A and N, we have:

$$ F(u^{I}_5) = x \quad (I7)$$

\textbf{Case 3.I.1.2.1.1}  $0 \leq |2a - 1| < \varepsilon$

If $a - (1 - a) < \varepsilon$, then voters in profile $u^{I}_4$ with preferences $\begin{pmatrix} x \\ y \\ z \end{pmatrix}$ would form a coalition of size $2a - 1$, and misreport their preferences as $\begin{pmatrix} z\\ y \\ x \end{pmatrix}$.  This would create profile $u^{I}_5$ and evaluation $(I6)$, giving improved result $x$.  This would contradict WSP, meaning that \textbf{Case 3.I.1.2.1.1} fails to hold.  

\textbf{Case 3.I.1.2.1.n+1}  $n\varepsilon \leq |2a - 1| < (n+1)\varepsilon$

Again, we have:

$$ F(u^{I}_4) = z \quad (I6)$$

$$ F(u^{I}_5) = x \quad (I6)$$

Now, consider the following profile:

$$u^{I}_{(j)} = \begin{tabular}{ c c }
a - jk &  1 - a + jk\\
\hline
 x  & z \\ 
y  & y \\  
z  & x 
\end{tabular}$$

Here, $j$ iterates from $1$ to $n$, and $k = \frac{2a - 1}{n + 1}$.  Note that the definition of $u^{I}_{(j)}$ is local to this case.  

We first consider the evaluation $F(u^{I}_{(1)})$.  Suppose that $F(u^{I}_{(1)}) \neq z$.  But then, voters in profile $u^{I}_4$ with preferences $\begin{pmatrix} x \\ y \\ z \end{pmatrix}$ would form a coalition of size $k$, and misreport their preferences as $\begin{pmatrix} z\\ y \\ x \end{pmatrix}$.  This would induce profile $u^{I}_{(1)}$ and an improved result over $z$, contradicting WSP.  So, we have that $F(u^{I}_{(1)}) = z$. 

Now, assume for a general $j$ that $F(u^{I}_{(j)}) = z$.  We can show that $F(u^{I}_{(j+1)}) = z$.  Suppose instead that $F(u^{I}_{(j+1)}) \neq z$.  But then, voters in profile $u^{I}_{(j)}$ with preferences $\begin{pmatrix} x \\ y \\ z \end{pmatrix}$ would form a coalition of size $k$, and misreport their preferences as $\begin{pmatrix} z\\ y \\ x \end{pmatrix}$.  This would induce profile $u^{I}_{(j+1)}$ and an improved result over $z$, contradicting WSP.  So, in general, $F(u^{I}_{(j+1)}) = z$.  In particular, this is true for the following profile:

$$u^{I}_{(n)} = \begin{tabular}{ c c }
a - nk &  1 - a + nk\\
\hline
 x  & z \\ 
y  & y \\  
z  & x 
\end{tabular} =  \begin{tabular}{ c c }
1 - a + k & a - k\\
\hline
 x  & z \\ 
y  & y \\  
z  & x 
\end{tabular} $$

But then, note that voters in profile $u^{I}_{(n)}$ with preferences $\begin{pmatrix} x \\ y \\ z \end{pmatrix}$ would form a coalition of size $k$, and misreport their preferences as $\begin{pmatrix} z\\ y \\ x \end{pmatrix}$.  This would induce profile $u^{I}_5$, and improved result $x$ over $z$, contradicting WSP.  This case fails to hold.  

This concludes \textbf{3.I.1.2} and \textbf{3.I.1}.  We next consider \textbf{3.I.2}.

\textbf{3.I.2} Suppose that the result of Condorcet $F^{C}(u_{.}^{I})$ is y. So, $a < \frac{1}{2}$, and $a + b + c > \frac{1}{2}$.

\textbf{3.I.2.1}
Suppose that:

$$F(u_{.}^{I}) = x \quad (I8)$$  

Consider the following profile:

$$u_{6}^{I} =  \begin{tabular}{ c c c c}
a  & b + c & 1 - a - b - c\\
\hline
 x  & y & z\\ 
y   & x & y\\  
z & z & x    
\end{tabular} $$

\textbf{3.I.2.1.1} Suppose that $$F(u_{6}^{I}) = y \quad (I9)$$

\textbf{Case 3.I.2.1.1.1} $0 \leq b < \varepsilon$

But then, note that voters in profile $u_{.}^{I}$ with preferences $\begin{pmatrix} y \\ z \\ x \end{pmatrix}$ will misreport as $\begin{pmatrix} y \\ x \\ z \end{pmatrix}$ with a coalition weight of $b < \varepsilon$.   This would induce profile $u_{6}^{I}$ and improved outcome $y$, violating WSP.  So, this fails to hold.

\textbf{Case 3.I.2.1.1.n+1} $n\varepsilon \leq b < (n+1)\varepsilon$

We are assuming that:

$$F(u_{6}^{I}) = y \quad (I9)$$

Now, consider the following profiles:

$$u^{I}_{(j)} = \begin{tabular}{ c c c c}
a & jk & b + c - jk & 1 - a - b - c\\
\hline
 x  & y & y & z\\ 
y  & z & x & y\\  
z  & x & z & x    
\end{tabular}$$

Here, $j$ iterates from $1$ to $n$, and $k = \frac{b}{n+1}$.  Also note that this definition of $u^{I}_{(j)}$ is local to this case.

Suppose that $F(u^{I}_{(1)}) \neq y$.  But then, voters in profile $u^{I}_{(1)}$ with preferences $\begin{pmatrix} y \\ z \\ x \end{pmatrix}$ will misreport as $\begin{pmatrix} y \\ x \\ z \end{pmatrix}$ with a coalition weight of $k < \varepsilon$.   This would induce profile $u_{6}^{I}$ and improved outcome $y$, violating WSP.  So, $F(u^{I}_{(1)}) = y$.  

Now, assume for a general $j$ that $F(u^{I}_{(j)}) = y$.  We can show that $F(u^{I}_{(j+1)}) = y$.  Suppose instead that $F(u^{I}_{(j+1)}) \neq y$.  But then, voters in profile $u^{I}_{(j+1)}$ with preferences $\begin{pmatrix} y \\ z \\ x \end{pmatrix}$ will misreport as $\begin{pmatrix} y \\ x \\ z \end{pmatrix}$ with a coalition weight of $k < \varepsilon$.   This would induce profile $u^{I}_{(j)}$ and improved outcome $y$, violating WSP.  So, $F(u^{I}_{(j+1)}) = y$.  

In particular, this is true for:

$$u^{I}_{(n)} = \begin{tabular}{ c c c c}
a & nk & b + c - nk & 1 - a - b - c\\
\hline
 x  & y & y & z\\ 
y  & z & x & y\\  
z  & x & z & x    
\end{tabular} =  \begin{tabular}{ c c c c}
a & b - k & b + c + k & 1 - a - b - c\\
\hline
 x  & y & y & z\\ 
y  & z & x & y\\  
z  & x & z & x    
\end{tabular}$$

So, voters in profile $u^{I}_.$ with preferences $\begin{pmatrix} y \\ z \\ x \end{pmatrix}$ will misreport as $\begin{pmatrix} y \\ x \\ z \end{pmatrix}$ with a coalition weight of $k < \varepsilon$.   This would induce profile $u^{I}_{(n)}$ and improved outcome $y$, violating WSP.  So, this case fails to hold.

\textbf{3.I.2.1.2} Next, suppose that $$F(u_{6}^{I}) = z \quad (I10)$$

\textbf{Case 3.I.2.1.2.1}  $0 \leq b < \varepsilon$

But then, voters in profile $u_{6}^{I}$ with preferences $\begin{pmatrix} y \\ x \\ z \end{pmatrix}$ will misreport as $\begin{pmatrix} y \\ z \\ x \end{pmatrix}$ with a weight of $b < \varepsilon$.  This will induce profile $u^{I}_.$ and evaluation $(I8)$, achieving improved result $x$ over $z$.  This would contradict WSP, so this case fails to hold.

\textbf{Case 3.I.2.1.2.n+1}  $n\varepsilon \leq b < (n+1)\varepsilon$

By assumption for this step, we have:

$$F(u_{.}^{I}) = x \quad (I8)$$

$$F(u_{6}^{I}) = z \quad (I10)$$

Consider the following profile:

$$u^{I}_{(j)} = \begin{tabular}{ c c c c}
a & jk & b + c - jk & 1 - a - b - c\\
\hline
 x  & y & y & z\\ 
y  & z & x & y\\  
z  & x & z & x \end{tabular}  $$ 

Here, $j$ iterates from $1$ to $n$, and $k = \frac{b}{n+1}$.  Note that the definition of $u^{I}_{(j)}$ is local to this case.

Suppose that $F(u^{I}_{(1)}) \neq z$.  But then, note that voters in profile $u_{6}^{I}$ with preferences $\begin{pmatrix} y \\ x \\ z \end{pmatrix}$ will misreport as $\begin{pmatrix} y \\ z \\ x \end{pmatrix}$ with a weight of $k < \varepsilon$.  This will induce profile $u^{I}_{(1)}$ and an improved result over $z$, contradicting WSP.  So, $F(u^{I}_{(1)}) = z$.  

Now, assume for a general $j$ that $F(u^{I}_{(j)}) = z$.  We can likewise show that $F(u^{I}_{(j+ 1)}) = z$.  Suppose instead that $F(u^{I}_{(j+ 1)}) \neq z$.  But then, voters in profile $u^{I}_{(j)}$ with preferences  $\begin{pmatrix} y \\ x \\ z \end{pmatrix}$ will misreport as $\begin{pmatrix} y \\ z \\ x \end{pmatrix}$ with a weight of $k < \varepsilon$.  This will induce profile $u^{I}_{(j+1)}$ and an improved result over $z$, contradicting WSP.  So, in general, $F(u^{I}_{(j+1)}) = z$.  

In particular, this is true for the following profile:

$$u^{I}_{(n)} = \begin{tabular}{ c c c c}
a & nk & b + c - nk & 1 - a - b - c\\
\hline
 x  & y & y & z\\ 
y  & z & x & y\\  
z  & x & z & x \end{tabular} = \begin{tabular}{ c c c c}
a & b - k & c + k & 1 - a - b - c\\
\hline
 x  & y & y & z\\ 
y  & z & x & y\\  
z  & x & z & x \end{tabular} $$ 

But then, note that voters in profile $u^{I}_{(n)}$ with preferences $\begin{pmatrix} y \\ x \\ z \end{pmatrix}$ will misreport as $\begin{pmatrix} y \\ z \\ x \end{pmatrix}$ with a weight of $k < \varepsilon$.  This will induce profile $u^{I}_.$ and evaluation $(I8)$, for improved result $x$ over $z$, contradicting WSP.  So, this case fails to hold.

\textbf{3.I.2.1.3} So, instead, it must be the case that  $$F(u_{6}^{I}) = x \quad (I11)$$

Now, we consider the following profile:

$$ u_{7}^{I} = \begin{tabular}{ c c}
a & 1 - a\\
\hline
 x  & y \\ 
y & x \\  
z  & z \end{tabular} $$

Note that $z$ is Pareto dominated here, so cannot be the result of the evaluation $F(u_{7}^{I})$.

\textbf{3.I.2.1.3.1}  Suppose that $$F(u_{7}^{I}) = y \quad (I12)$$

\textbf{Case 3.I.2.1.3.1.1} $0 \leq 1 - a - b - c < \varepsilon$

In this case, voters in profile $u_{6}^{I}$ with preferences $\begin{pmatrix} z \\ y \\ x \end{pmatrix}$ will misreport as $\begin{pmatrix} y \\ x \\ z \end{pmatrix}$ with a weight of $1 - a - b - c < \varepsilon$.  This would induce profile $u_{7}^{I}$ and improved result $y$ over $x$, contradicting WSP.  So, \textbf{Case 3.I.2.1.3.1.1} fails to hold.

\textbf{Case 3.I.2.1.3.1.n+1} $n\varepsilon \leq 1 - a - b - c < (n+1)\varepsilon$

Again, for \textbf{3.I.2.1.3}, we have that:

 $$F(u_{6}^{I}) = x \quad (I11)$$
 
 We are assuming that:
 
 $$F(u_{7}^{I}) = y \quad (I12)$$

Now, consider the following profile:

$$u^{I}_{(j)} = \begin{tabular}{ c c c}
a &  b + c +jk & 1 - a - b - c - jk\\
\hline
 x  & y & z\\ 
y  & x & y\\  
z  & z & x \end{tabular}$$

Here, $j$ iterates from $1$ to $n$, and $k = \frac{1 - a - b - c}{n+1}$.  Note that this definition of $u^{I}_{(j)}$ is local to this case.

We first consider $u^{I}_{(1)}$.  Suppose that $F(u^{I}_{(1)}) \neq x$.  But then, voters in profile $u_{6}^{I}$ with preferences $\begin{pmatrix} z \\ y \\ x \end{pmatrix}$ will misreport as $\begin{pmatrix} y \\ x \\ z \end{pmatrix}$ with a weight of $k < \varepsilon$.  This would induce profile $u^{I}_{(1)}$ and an improvement over $x$, contradicting WSP.  So, $F(u^{I}_{(1)}) = x$.

Now, assume that for some $j$ that $F(u^{I}_{(j)}) = x$.  We can show that $F(u^{I}_{(j+1)}) = x$. Suppose instead that $F(u^{I}_{(j+1)}) \neq x$. But then, note that voters in profile $u^{I}_{(j)}$ with preferences $\begin{pmatrix} z \\ y \\ x \end{pmatrix}$ will misreport as $\begin{pmatrix} y \\ x \\ z \end{pmatrix}$ with a weight of $k < \varepsilon$.  This would induce profile $u^{I}_{(j+1)}$ and an improvement over $x$, contradicting WSP.  So, in general, $F(u^{I}_{(j+1)}) = x$.

This is true in particular for the following profile:

$$u^{I}_{(n)} = \begin{tabular}{ c c c}
a &  b + c +nk & 1 - a - b - c - nk\\
\hline
 x  & y & z\\ 
y  & x & y\\  
z  & z & x \end{tabular} = \begin{tabular}{ c c c}
a & 1 - a - k & k \\
\hline
 x  & y & z\\ 
y  & x & y\\  
z  & z & x \end{tabular} $$

But then, note that voters in profile $u^{I}_{(n)}$ with preferences $\begin{pmatrix} z \\ y \\ x \end{pmatrix}$ will misreport as $\begin{pmatrix} y \\ x \\ z \end{pmatrix}$ with a weight of $k < \varepsilon$.  This will induce profile $u^I_7$, achieving improved outcome $y$ over $x$, contradicting WSP.  So, \textbf{3.I.2.1.3.1} fails to hold.

\textbf{3.I.2.1.3.2}  Instead, it must be true that $$F(u_{7}^{I}) = x \quad (I13)$$

Recall that:
$$ u_{7}^{I} = \begin{tabular}{ c c}
a & 1 - a\\
\hline
 x  & y \\ 
y & x \\  
z  & z \end{tabular} $$

By assumption for \textbf{3.I.2}, the Condorcet winner $F^C(u^{I}_.) = y$, implying that $a < \frac{1}{2}$, and $1 - a > \frac{1}{2}$.

Consider the following relabeling of the weights on profile $u_{7}^{I}$, and apply permutation $\sigma^*$, where $\sigma^*(x) = y$, $\sigma^*(y) = x$, and $\sigma^*(z) = z$.

$$ u_{7}^{I} = \begin{tabular}{ c c}
A & 1 - A \\
\hline
  y & x\\ 
x & y\\  
z & z \end{tabular} $$

$$F(u_{7}^{I}) = x \quad (I13)$$

(Here, $A = 1 - a > \frac{1}{2}$).

$$ u_{8}^{I} = \begin{tabular}{ c c}
A & 1 - A \\
\hline
 x  & y \\ 
y & x \\  
z  & z \end{tabular} $$

By N and A:

$$F(u_{8}^{I}) = y \quad (I13)$$

Now, note that we can derive a similar contradiction to WSP as we did in \textbf{3.I.1.1.1}.  So, this case fails to hold.

\textbf{3.I.2.2}  Suppose that $F(u^{I}_.) = z$.  But, note the symmetry between $x$ and $z$, meaning that the impossibility of \textbf{3.I.2.1} also implies the impossibility of \textbf{3.I.2.2}.

\textbf{3.I.3}  Suppose that the result of Condorcet $F^{C}(u_{.}^{I})$ is $z$.  But, we note that $x$ and $z$ are symmetric in this domain, meaning that this case has been ruled out by consideration of \textbf{3.I.1}.  So, we move on to the next domain.

\textbf{STEP 3, Domain II (3.II)}

Consider the next following rich domain:

$$U^{II} = \Bigg\{ \begin{tabular}{ c c c c}
 x  & y & y \\ 
y  & z & x\\  
z  & x & z    
\end{tabular} \Bigg\} $$

Now we assume that $F$ is some voting rule that satisfies P, A, N, and WSP on $U^{II}$.  We want to show that $F$ must equal $F^{C}$.  So, we proceed by contradiction.  Fix $\varepsilon > 0$, and fix $a$.  

We consider the following generic profile, where $F(u^{II}_.) \neq F^{C}(u^{II}_.)$.

$$ u_.^{II} = \begin{tabular}{ c c c }
a & b & 1 - a - b\\
\hline
 x & y & y \\ 
y & z & x \\  
z & x & z    
\end{tabular} $$

We see that $z$ is Pareto-dominated by $y$ - so, the result of voting rules $F$ and $F^{C}$ on $u^{I}_.$ cannot be $z$.

\textbf{3.II.1} Suppose that the Condorcet winner $F^{C}(u^{II}_{.}) = x$.

This means that $a > \frac{1}{2}$ by definition of the Condorcet method.  As we are assuming that $F \neq F^{C}$, we have that:

$$F(u^{II}_{.}) = y \quad (I14)$$

Consider the following profile:

$$ u^{II}_1 =  \begin{tabular}{ c c }
a &  1 - a\\
\hline
 x  & y \\ 
y  & x \\  
z  & z 
\end{tabular} $$

\textbf{3.II.1.0}  We want to show that for a general $b$, $F(u^{II}_1)= y$.

\textbf{Case 3.II.1.0.1} $0 \leq b < \varepsilon$

Suppose that $F(u^{II}_1)= x$.  But then, voters in profile $u^{II}_1$ with preferences $\begin{pmatrix} y \\ x \\ z \end{pmatrix}$ would misreport as $\begin{pmatrix} y \\ z \\ x \end{pmatrix}$ with a coalition size of $b < \varepsilon$.  This would induce profile $u^{II}_.$ and evaluation $(I14)$, and would yield improved result $y$, contradicting WSP.  So, for this case, it must be true that $F(u^{II}_1)= x$.

\textbf{Case 3.II.1.0.2} $\varepsilon \leq b < 2\varepsilon$

By assumption for \textbf{3.II.1}, we have that:

$$F(u^{II}_{.}) = y \quad (I14)$$

Now, consider the following profile:

$$ u^{II}_2 = \begin{tabular}{ c c c }
a & b - k & 1 - a - b + k\\
\hline
 x & y & y \\ 
y & z & x \\  
z & x & z    
\end{tabular} $$

Here, $k = \frac{b}{2}$.

For $F$ to satisfy WSP, then $F(u^{II}_2) = y$.  Suppose instead that $F(u^{II}_2) = x$.  But then, voters in profile $u^{II}_2$ with preferences $\begin{pmatrix} y \\ x \\ z \end{pmatrix}$ would form a coalition of size $k < \varepsilon$, and misreport their preferences as $\begin{pmatrix} y \\ z \\ x \end{pmatrix}$.  This would induce profile $u^{II}_.$ and evaluation $(I14)$, achieving improved outcome $y$, contradicting WSP.  So, we know that $F(u^{II}_2) = y$. 

Also note that:

$$ u^{II}_2 = \begin{tabular}{ c c c }
a & b - k & 1 - a - b + k\\
\hline
 x & y & y \\ 
y & z & x \\  
z & x & z    
\end{tabular} = \begin{tabular}{ c c c }
a & k & 1 - a - k\\
\hline
 x & y & y \\ 
y & z & x \\  
z & x & z    
\end{tabular} $$

This will imply that $F(u^{II}_1) = y$.  Suppose instead that $F(u^{II}_1) = x$.  But then, voters in profile $u^{II}_1$ with preferences $\begin{pmatrix} y \\ x \\ z \end{pmatrix}$ would form a coalition of size $k < \varepsilon$, and misreport their preferences as $\begin{pmatrix} y \\ z \\ x \end{pmatrix}$.  This would induce profile $u^{II}_2$ and improved result $y$, contradicting WSP.  So, we have for this case that $F(u^{II}_1) = x$.

We can also see that $F(u^{II}_1) = y$ by applying \textbf{3.II.1.0.1}.  Note that $k < \varepsilon$.  So, profile $u^{II}_2$ is just a case of the profiles that we considered in \textbf{Case 3.II.1.0.1} to show that $F(u^{II}_1) = y$.  Because we have shown that $F(u^{II}_2) = y$, then this in turn implies that $F(u^{II}_1) = y$.

Now, for the inductive hypothesis, we assume that all cases up to \textbf{Case 3.II.1.0.n} hold.  In other words, given some profile:

$$ u^{*} = \begin{tabular}{ c c c }
$a$ & $b^{*}$ & $1 - a - b^{*}$\\
\hline
 x & y & y \\ 
y & z & x \\  
z & x & z    
\end{tabular} $$

where the following conditions are true:

$$ F(u^{*}) = y$$

$$(n- 1)\varepsilon \leq b^* < n\varepsilon \quad (W6)$$

then this implies that $F(u^{II}_1) = y$.

Now, we consider the following case:

\textbf{Case 3.II.1.0.n+1} $n\varepsilon \leq b < (n+1)\varepsilon$

Again, by assumption for \textbf{3.II.1}, we have that:

$$F(u^{II}_{.}) = y \quad (I14)$$

Now, consider the following profile:

$$ u^{II}_k = \begin{tabular}{ c c c }
a & b - k & 1 - a - b + k \\
\hline
 x & y & y \\ 
y & z & x \\  
z & x & z    
\end{tabular} $$

(Here, $k = \frac{b}{n+1}$.  Note that this definition of $k$ is local to this step).

Suppose that $F(u^{II}_k) = x$.  But then, voters in profile $u^{II}_k$ with preferences  $\begin{pmatrix} y \\ x \\ z \end{pmatrix}$ would form a coalition of size $k < \varepsilon$, and misreport their preferences as $\begin{pmatrix} y \\ z \\ x \end{pmatrix}$.  This would induce profile $u^{II}_{.}$ and evaluation $(I14)$, resulting in improved outcome $y$.  This would contradict WSP, meaning that $F(u^{II}_k) = y$.

But then, consider the weights in profile $u^{II}_k$.  Note that: $$b - k = \frac{nb}{n + 1}$$

And:
$$(n - 1)\varepsilon \leq \frac{nb}{n + 1} < n\varepsilon$$

These weights fulfill inequality $(W6)$.  So, given that $F(u^{II}_k) = y$, by the inductive hypothesis, we can conclude that in this case, $F(u^{II}_1) = y$, as we wanted.  
Thus, in general, for $F$ to satisfy WSP, it must be true that:

$$ F(u^{II}_1) = y \quad (I15)$$

\textbf{3.II.1.1} Now, we will show that evaluation $(I15)$ results in a violation of WSP.

Apply permutation $\sigma^*$ to profile $u^{II}_1$, where $\sigma^*(x) = y, \sigma^*(y) = x$, and $\sigma^*(z) = z$.

This gives us the following profile and evaluation:

$$  u^{II}_3 =  \begin{tabular}{ c c }
a &  1 - a\\
\hline
 y  & x \\ 
x  & y \\  
z  & z 
\end{tabular}$$

$$ F(u^{II}_3) = x \quad (I16)$$

\textbf{Case 3.II.1.1.1}  $0 \leq |2a - 1| < \varepsilon$

If $a - (1 - a) < \varepsilon$, then voters in profile $u^{II}_3$ with preferences $\begin{pmatrix} y \\ x \\ z \end{pmatrix}$ would form a coalition of size $2a - 1$, and misreport their preferences as $\begin{pmatrix} x\\ y \\ z \end{pmatrix}$.  This would create profile $u^{II}_1$ and evaluation $(I15)$, giving improved result $y$.  

This would contradict WSP, meaning that on this range of $a$ and $b$, our original hypothesis that $F \neq F^{C}$ cannot be true.  So, this tells us that when $a^{*}$ fulfills the inequality, $F$ agrees with $F^C$:

$$ \frac{1}{2} < a^* < \frac{1}{2} + \frac{\varepsilon}{2} \quad (W7)$$

Then:

$$
F \left(
\begin{tabular}{ c c c }
$a^*$ & $1 - a^*$\\
\hline
 x & y \\ 
y & x \\  
z & z    
\end{tabular} \right) = x
$$

\textbf{Case 3.II.1.1.2}  $\varepsilon \leq |2a - 1| < 2\varepsilon$ (i.e., $\frac{\varepsilon}{2} + \frac{1}{2} < a \leq \varepsilon + \frac{1}{2}$)

Note that $a = \frac{1}{2} + \varepsilon - \delta$ for some $0 < \delta \leq \frac{\varepsilon}{2}$.

Again, we have:

$$ F(u^{II}_1) = y \quad (I15)$$

Now, consider the following profile:

$$u^{II}_4 = \begin{tabular}{ c c }
a - k &  1 - a + k\\
\hline
 x  & y \\ 
y  & x \\  
z  & z 
\end{tabular}$$

Where:  $$k = \frac{a - \frac{1}{2}}{2} = \frac{\varepsilon - \delta}{2} < \varepsilon$$

Note that $a - k = \frac{1}{2} + \frac{\varepsilon}{2} - \frac{\delta}{2} < \frac{1}{2} + \frac{\varepsilon}{2}$

Additionally, note that:

$$ \frac{\delta}{2} < \frac{\varepsilon}{4}$$

$$ \frac{\varepsilon}{4} - \frac{\delta}{2} > 0 $$

So, for the weight on preferences $\begin{pmatrix} x \\ y \\ z \end{pmatrix}$ in profile $u^{II}_4$:

$$a - k = \frac{1}{2} + \frac{\varepsilon}{2} - \frac{\delta}{2} = \frac{1}{2} + \frac{\varepsilon}{4} + \frac{\varepsilon}{4} - \frac{\delta}{2} > \frac{1}{2} + \frac{\varepsilon}{4} $$

This means that:

$$ \frac{1}{2} < a - k < \frac{1}{2} + \frac{\varepsilon}{2}$$

So, this weight is within the range $(W7)$ where $F = F^{C}$ from \textbf{Case 3.II.1.1.1}.  So, $F(u^{II}_4) = x$.

But then, note that voters in profile $u^{II}_1$ with preferences $\begin{pmatrix} x\\ y \\ z \end{pmatrix}$ would misreport as $\begin{pmatrix} y\\ x \\ z \end{pmatrix}$ with a weight of $k$ to induce profile $u^{II}_4$ and improved result $x$.  This would contradict WSP, meaning that on this range of $(a, b)$, $F = F^{C}$.

In other words, for a profile, where:

$$ \frac{1}{2} < a^* < \frac{1}{2} + \varepsilon $$

We can say that:

$$
F \left(
\begin{tabular}{ c c }
$a^*$ &  $1 - a^*$\\
\hline
 x  & z \\ 
y  & y \\  
z  & x    
\end{tabular} \right) = x
$$

A similar logic will apply for all $a > \frac{1}{2}$.  So, we see that for \textbf{3.II.1}, there is no voting rule $F \neq F^C$ that satisfies the properties that we want.

\textbf{3.II.2}  Now suppose that the Condorcet winner $F^{C}(u^{II}_{.}) = y$.

This means that $a < \frac{1}{2}$. 

We are assuming that $F \neq F^{C}$, so, we have that:

$$F(u^{II}_{.}) = x \quad (I17)$$

Consider the profile:
$$ u^{II}_5 =  \begin{tabular}{ c c }
a &  1 - a\\
\hline
 x  & y \\ 
y  & x \\  
z  & z 
\end{tabular}   $$

We want to show that $F(u^{II}_5) = x$.

\smallskip

\textbf{Case 3.II.2.1} $0 \leq b < \varepsilon$

Assume that the following is true, and proceed by contradiction.

$$F \left(
\begin{tabular}{ c c }
a &  1 - a\\
\hline
 x  & y \\ 
y  & x \\  
z  & z    
\end{tabular}\right) = F(u^{II}_5) = y \quad 
$$

Then, those with true preference ranking $\begin{pmatrix} y \\ z \\ x \end{pmatrix}$ in profile $u^{II}_{.}$ would form a coalition of size $b < \varepsilon$ and misreport their true preferences as $\begin{pmatrix} y \\ x \\ z \end{pmatrix}$, which would form profile $u^{II}_5$, inducing result $y$, meaning that this would be a profitable deviation.  This contradicts WSP, and tells us for the case where b $< \varepsilon$, then $F(u^{II}_5) = x$.

\smallskip

\textbf{Case 3.II.2.2} $\varepsilon \leq b < 2\varepsilon$

Consider the profile:
$$ u^{II}_6= \begin{tabular}{ c c c }
a & m & 1 - a - m\\
\hline
 x & y & y \\ 
y & z & x \\  
z & x & z    
\end{tabular}$$

(where $m = \frac{b}{2} $).

Then, $F(u^{II}_6)$ must equal $x$, because otherwise, from profile $u^{II}_{.}$, a coalition of size $b - m$ would deviate from $\begin{pmatrix} y \\ z \\ x \end{pmatrix}$ and misreport as $\begin{pmatrix} y \\ x \\ z \end{pmatrix}$ to form profile $u^{II}_6$, and achieve result $y$.  This would contradict WSP (because the coalition size $b - m < \varepsilon$), so $F(u^{II}_6) = x$.  Also note that $m < \varepsilon$. 

Then, we consider $F(u^{II}_5)$.  If $F(u^{II}_5) = y$, then all voters in $u^{II}_6$ with preferences $\begin{pmatrix} y \\ z \\ x \end{pmatrix}$ would misreport as $\begin{pmatrix} y \\ x \\ z \end{pmatrix}$ to form profile $u^{II}_5$.  So, this would contradict WSP, meaning that $F(u^{II}_6) = x$.

A similar logic applies for all subsequent $b$ values.  So, in general, for any $a < \frac{1}{2}$ it must be true that:

$$ F \left(
\begin{tabular}{ c c }
a &  1 - a\\
\hline
 x  & y \\ 
y  & x \\  
z  & z    
\end{tabular}\right) = x \quad $$

If we rename $1 - a$ and call it $A$, where $A > \frac{1}{2}$, then we have:

$$ F \left(
\begin{tabular}{ c c }
A &  1 - A\\
\hline
 y  & x \\ 
x  & y \\  
z  & z    
\end{tabular}\right) = x \quad $$

Then, we can apply the permutation $\sigma^{*}$, where $\sigma^{*}(x) = y, \sigma^{*}(y) = x, \sigma^{*}(z) = z$.  By N and A, we have:

$$ F \left(
\begin{tabular}{ c c }
A &  1 - A\\
\hline
 x  & y \\ 
y  & x \\  
z  & z    
\end{tabular}\right) = y \quad $$

This is the same pair of profiles that we had in \textbf{3.II.1}, so the same results follow, leading to a contradiction that tells us that $F = F^{C}$.  So, this shows that for \textbf{Domain II}, there is no voting rule $F$ other than the Condorcet method that satisfies the axioms on this domain.

\textbf{STEP 3, Domain III}

We will now consider the final domain:

$$U^{III} = \Bigg\{ \begin{tabular}{ c c c c}
 x  & x & y & y\\ 
y  & z & z & x\\  
z  & y & x & z    
\end{tabular} \Bigg\}  $$

Now we assume that $F$ is some voting rule that satisfies the P, A, N, and WSP, where $F \neq F^{C}$.  

So, $F$ must differ from $F^{C}$ for some profile $u_.^{III}$:

$$ u_.^{III} = \begin{tabular}{ c c c c}
a & b & c & 1 - a - b - c\\
\hline
 x  & x & y & y\\ 
y  & z & z & x\\  
z  & y & x & z     
\end{tabular} $$

Note that the Condorcet winner on profile $u_.^{III}$ can never be $z$.  Also note that $x$ and $y$ are symmetric in the profile.  
So, we assume without loss of generality that the Condorcet winner of $u_.^{III}$ is $x$.  This means that:

$$ a + b > 1 - a - b$$

\textbf{3.III.1}  Suppose that $$F(u_.^{III}) = y \quad (I18)$$

Consider the following profile:

$$ u^{III}_1 = \begin{tabular}{ c c c}
a + b & c & 1 - a - b - c\\
\hline
 x  & y & y\\ 
y  & z & x\\  
z  & x & z     
\end{tabular}
$$ 

\textbf{3.III.1.1} Suppose that $$F(u^{III}_1) = x$$

\textbf{Case 3.III.1.1.1} $0 \leq b < \varepsilon$

In this case, voters in profile $u_.^{III}$ with preferences $\begin{pmatrix} x \\ z \\ y \end{pmatrix}$ will misreport as $\begin{pmatrix} x \\ y \\ z \end{pmatrix}$ with a coalition size of $b$, inducing profile $u^{III}_1$ and improved result $x$.  This would contradict WSP, so \textbf{Case 3.III.1.1.1} fails to hold.

\textbf{Case 3.III.1.1.n+1} $n\varepsilon \leq b < (n+1)\varepsilon$

Consider the following profile:

$$ u^{III}_{(j)} = \begin{tabular}{ c c c c}
a + b - jk & jk & c & 1 - a - b - c\\
\hline
 x  & x & y & y\\ 
y  & z & z & x\\  
z  & y & x & z     
\end{tabular}
$$ 

Here, $k = \frac{b}{n+1}$, and $j$ iterates from $1$ to $n$.  Note that the definition of $u^{III}_{(j)}$ is local to this case.

We first consider $F(u^{III}_{(1)})$.  Suppose that $F(u^{III}_{(1)}) \neq x$.  But then, voters in profile $u^{III}_{(1)}$ with preferences $\begin{pmatrix} x \\ z \\ y \end{pmatrix}$ will misreport as $\begin{pmatrix} x \\ y \\ z \end{pmatrix}$ with a coalition size of $k < \varepsilon$, inducing profile $u^{III}_1$ and improved result $x$.  This would contradict WSP, so we know that $F(u^{III}_{(1)}) = x$.

Now, assume for a general $j$ that $F(u^{III}_{(j)}) = x$.  We can show that it follows that $F(u^{III}_{(j+1)}) = x$.  Suppose instead that $F(u^{III}_{(j+1)}) \neq x$.  But then, voters in profile $u^{III}_{(j+1)}$ with preferences $\begin{pmatrix} x \\ z \\ y \end{pmatrix}$ will misreport as $\begin{pmatrix} x \\ y \\ z \end{pmatrix}$ with a coalition size of $k < \varepsilon$, inducing profile $u^{III}_{(j)}$, and improved outcome $x$.  This would contradict WSP, meaning that $F(u^{III}_{(j+1)}) = x$.  This is true for any general $j$.  In particular, this is true for:

$$ u^{III}_{(n)} = \begin{tabular}{ c c c c}
a + b - nk & nk & c & 1 - a - b - c\\
\hline
 x  & x & y & y\\ 
y  & z & z & x\\  
z  & y & x & z     
\end{tabular}
 = \begin{tabular}{ c c c c}
a + k & b - k & c & 1 - a - b - c\\
\hline
 x  & x & y & y\\ 
y  & z & z & x\\  
z  & y & x & z     
\end{tabular}$$ 

But now, note that voters in profile $u_.^{III}$ with preferences $\begin{pmatrix} x \\ z \\ y \end{pmatrix}$ will misreport as $\begin{pmatrix} x \\ y \\ z \end{pmatrix}$ with a coalition size of $k$, inducing profile $u^{III}_{(n)}$ and improved outcome $x$ over $y$.  So, \textbf{3.III.1.1} fails to hold. 

\textbf{3.III.1.2} Now suppose that $$F(u^{III}_1) = z $$

\textbf{Case 3.III.1.2.1} 

In this case, some voters in profile $u^{III}_1$ with preferences $\begin{pmatrix} x \\ y \\ z \end{pmatrix}$ will misreport as $\begin{pmatrix} x \\ z \\ y \end{pmatrix}$ with a coalition size of $b$, inducing profile $u^{III}_.$ and evaluation $(I18)$, giving improved result $y$ over $z$.  This would contradict WSP, so \textbf{Case 3.III.1.2.1} fails to hold.

\textbf{Case 3.III.1.2.n+1} $n\varepsilon \leq b < (n+1)\varepsilon$

Consider the following profile:

$$ u^{III}_{(j)} = \begin{tabular}{ c c c c}
a + b - jk  & jk & c & 1 - a - b - c\\
\hline
 x  & x & y & y\\ 
y  & z & z & x\\  
z  & y & x & z     
\end{tabular}
$$ 

Here, $k = \frac{b}{n+1}$, and $j$ iterates from $1$ to $n$.  Note that the definition of $u^{III}_{(j)}$ is local to this case.

We first consider $F(u^{III}_{(1)})$.  Suppose that $F(u^{III}_{(1)}) \neq z$.  But then, voters in profile $u^{III}_1$ with preferences $\begin{pmatrix} x \\ y \\ z\end{pmatrix}$ will misreport as $\begin{pmatrix} x \\ z \\ y \end{pmatrix}$ with a coalition size of $k < \varepsilon$, inducing profile $u^{III}_{(1)}$ and improved result over $z$.  This would contradict WSP, so we know that $F(u^{III}_{(1)}) = z$.

Now, assume for a general $j$ that $F(u^{III}_{(j)}) = z$.  We can show that it follows that $F(u^{III}_{(j+1)}) = z$.  Suppose instead that $F(u^{III}_{(j+1)}) \neq z$.  But then, voters in profile $u^{III}_{(j)}$ with preferences $\begin{pmatrix} x \\ y \\ z \end{pmatrix}$ will misreport as $\begin{pmatrix} x \\ z \\ y \end{pmatrix}$ with a coalition size of $k < \varepsilon$, inducing profile $u^{III}_{(j+1)}$, and improved outcome $x$.  This would contradict WSP, meaning that $F(u^{III}_{(j+1)}) = z$.  This is true for any general $j$.  In particular, this is true for:

$$ u^{III}_{(n)} = \begin{tabular}{ c c c c}
a + b - nk & nk & c & 1 - a - b - c\\
\hline
 x  & x & y & y\\ 
y  & z & z & x\\  
z  & y & x & z     
\end{tabular}
 = \begin{tabular}{ c c c c}
a + k & b - k & c & 1 - a - b - c\\
\hline
 x  & x & y & y\\ 
y  & z & z & x\\  
z  & y & x & z     
\end{tabular}$$ 

But now, note that voters in profile $u^{III}_{(n)}$ with preferences $\begin{pmatrix} x \\ y \\ z\end{pmatrix}$ will misreport as $\begin{pmatrix} x \\ z \\ y \end{pmatrix}$ with a coalition size of $k$, inducing profile $u^{III}_.$ and evaluation $(I18)$, yielding improved outcome $y$ over $z$.  So, \textbf{3.III.1.2} fails to hold. 

\textbf{3.III.1.3}  Instead, we have that: 

$$ F(u^{III}_1) = y$$

Where:

$$ u^{III}_1 = \begin{tabular}{ c c c}
a + b & c & 1 - a - b - c\\
\hline
 x  & y & y\\ 
y  & z & x\\  
z  & x & z     
\end{tabular}
$$

$$ a + b > 1 - a - b$$

But then, consider the following relabeling of the weights on profile $u^{III}_1$.

$$ u^{III}_1 = \begin{tabular}{ c c c}
a + b & c & 1 - a - b - c\\
\hline
 x  & y & y\\ 
y  & z & x\\  
z  & x & z     
\end{tabular}
 = \begin{tabular}{ c c c}
A & B & 1 - A - B\\
\hline
 x  & y & y\\ 
y  & z & x\\  
z  & x & z     
\end{tabular}$$ 

Where:  $a + b = A > \frac{1}{2}$.

But then, note that in \textbf{3.II.1}, we showed that for a generic profile $u^*$, and voting rule $F$ (satisfying P, A, N, and WSP), where:

$$ u^{*} = \begin{tabular}{ c c c}
$a^{*}$ & $b^{*}$ & $1 - a^* - b^*$\\
\hline
 x  & y & y\\ 
y  & z & x\\  
z  & x & z     
\end{tabular} $$

$$F(u^{*}) = y$$

$$ a^* > \frac{1}{2}$$

Then:

$$ F \left( \begin{tabular}{ c c c}
$a^{*}$  & $1 - a^*$ \\
\hline
 x  & y\\ 
y   & x\\  
z   & z     
\end{tabular} \right) = y$$

But, as we showed in \textbf{3.II.1}, this led to a contradiction with WSP.  

So, \textbf{3.III.1} is impossible.

\textbf{3.III.2} Instead we suppose that $$F(u_.^{III}) = y \quad (I19)$$

Again, we consider the following profile:

$$ u^{III}_1 = \begin{tabular}{ c c c}
a + b & c & 1 - a - b - c\\
\hline
 x  & y & y\\ 
y  & z & x\\  
z  & x & z     
\end{tabular}
$$ 

\textbf{3.III.2.1} Suppose that $$F(u^{III}_1) = x$$

\textbf{Case 3.III.2.1.1} 

But then, voters in profile $u_.^{III}$ with preferences $\begin{pmatrix} x \\ z \\ y \end{pmatrix}$ will misreport as $\begin{pmatrix} x \\ y \\ z \end{pmatrix}$ with a coalition size of $b$, inducing profile $u^{III}_1$ and improved result $x$.  This would contradict WSP, so \textbf{Case 3.III.2.1.1} fails to hold.

\textbf{Case 3.III.1.1.n+1} $n\varepsilon \leq b < (n+1)\varepsilon$

Consider the following profile:

$$ u^{III}_{(j)} = \begin{tabular}{ c c c c}
a + b - jk & jk & c & 1 - a - b - c\\
\hline
 x  & x & y & y\\ 
y  & z & z & x\\  
z  & y & x & z     
\end{tabular}
$$ 

Here, $k = \frac{b}{n+1}$, and $j$ iterates from $1$ to $n$.  Note that the definition of $u^{III}_{(j)}$ is local to this case.

We first consider $F(u^{III}_{(1)})$.  Suppose that $F(u^{III}_{(1)}) \neq x$.  But then, voters in profile $u^{III}_{(1)}$ with preferences $\begin{pmatrix} x \\ z \\ y \end{pmatrix}$ will misreport as $\begin{pmatrix} x \\ y \\ z \end{pmatrix}$ with a coalition size of $k < \varepsilon$, inducing profile $u^{III}_1$ and improved result $x$.  This would contradict WSP, so we know that $F(u^{III}_{(1)}) = x$.

Now, assume for a general $j$ that $F(u^{III}_{(j)}) = x$.  We can show that it follows that $F(u^{III}_{(j+1)}) = x$.  Suppose instead that $F(u^{III}_{(j+1)}) \neq x$.  But then, voters in profile $u^{III}_{(j+1)}$ with preferences $\begin{pmatrix} x \\ z \\ y \end{pmatrix}$ will misreport as $\begin{pmatrix} x \\ y \\ z \end{pmatrix}$ with a coalition size of $k < \varepsilon$, inducing profile $u^{III}_{(j)}$, and improved outcome $x$.  This would contradict WSP, meaning that $F(u^{III}_{(j+1)}) = x$.  This is true for any general $j$.  In particular, this is true for:

$$ u^{III}_{(n)} = \begin{tabular}{ c c c c}
a + b - nk & nk & c & 1 - a - b - c\\
\hline
 x  & x & y & y\\ 
y  & z & z & x\\  
z  & y & x & z     
\end{tabular}
 = \begin{tabular}{ c c c c}
a + k & b - k & c & 1 - a - b - c\\
\hline
 x  & x & y & y\\ 
y  & z & z & x\\  
z  & y & x & z     
\end{tabular}$$ 

But now, note that voters in profile $u_.^{III}$ with preferences $\begin{pmatrix} x \\ z \\ y \end{pmatrix}$ will misreport as $\begin{pmatrix} x \\ y \\ z \end{pmatrix}$ with a coalition size of $k$, inducing profile $u^{III}_{(n)}$ and improved outcome $x$ over $z$.  So, \textbf{3.III.2.1} fails to hold.

\textbf{3.III.2.2}  Suppose that $$F(u^{III}_1) = y$$

But, using the same logic as with \textbf{3.III.1.3}, this will fail to satisfy WSP.

\textbf{3.III.2.3}  So, we are left with: $$F(u^{III}_1) = z$$

Consider the following profile:

$$ u^{III}_2 = \begin{tabular}{ c c}
a + c & 1 - a - c\\
\hline
 y & y\\ 
 z & x\\  
x & z     
\end{tabular}
$$ 

We see that $y$ must be the winner (by property P).  

\textbf{Case 3.III.2.3.1} $0 \leq a + b < \varepsilon$

Here, voters in profile $u^{III}_1$ with preferences $\begin{pmatrix} x \\ y \\ z \end{pmatrix}$ will misreport as $\begin{pmatrix} y \\ z \\ x \end{pmatrix}$ with a weight of $a$, and will misreport as $\begin{pmatrix} y \\ x \\ z \end{pmatrix}$ with a weight of $b$.  This will create profile $u^{III}_2$, resulting in improved outcome $y$ over $z$.  This would contradict WSP, meaning that \textbf{Case 3.III.2.3.1} fails to hold.

\textbf{Case 3.III.2.3.n+1} $n\varepsilon \leq a + b < (n+1)\varepsilon$

Now, consider the following profile:

$$ u^{III}_{(j)} = \begin{tabular}{ c c c}
a + b - j(k + m) & c + jk & 1 - a - b - c + jm\\
\hline
 x   & y & y\\ 
y   & z & x\\  
z   & x & z     
\end{tabular}$$

Here, $k = \frac{a}{n+1}$ and $m = \frac{b}{n+1}$.  Also, $j$ iterates from $1$ to $n$.  Note that this definition of $u^{III}_{(j)}$ is local to this case.

We consider the evaluation $F(u^{III}_{(1)})$.  Suppose that $F(u^{III}_{(1)}) \neq z$.  But then, voters in profile $u^{III}_1$ with preferences $\begin{pmatrix} x \\ y \\ z \end{pmatrix}$ will misreport as $\begin{pmatrix} y \\ z \\ x \end{pmatrix}$ with a weight of $k$, and will misreport as $\begin{pmatrix} y \\ x \\ z \end{pmatrix}$ with a weight of $m$.  This would induce profile $u^{III}_{(1)}$ and would result in an improved outcome over $z$.  This would contradict WSP, meaning that $F(u^{III}_{(1)}) = z$.

Now, assume for some general $j$ that $F(u^{III}_{(j)}) = z$.  We can show that $F(u^{III}_{(j+1)}) = z$.  Suppose instead that $F(u^{III}_{(j + 1)}) \neq z$.  But then, voters in profile $u^{III}_{(j)}$ with preferences $\begin{pmatrix} x \\ y \\ z \end{pmatrix}$ will misreport as $\begin{pmatrix} y \\ z \\ x \end{pmatrix}$ with a weight of $k$, and will misreport as $\begin{pmatrix} y \\ x \\ z \end{pmatrix}$ with a weight of $m$.  This would induce profile $u^{III}_{(j+1)}$ and would result in an improved outcome over $z$.  So, for a general $j$, we have that $F(u^{III}_{(j+1)}) = z$.  In particular, this is true for the following profile:

$$ u^{III}_{(n)} = \begin{tabular}{ c c c}
a + b - n(k + m) & c + nk & 1 - a - b - c + nm\\
\hline
 x   & y & y\\ 
y   & z & x\\  
z   & x & z     
\end{tabular}$$ 

$$ = \begin{tabular}{ c c c}
(k + m) & a + c - k & 1 - a - c - m\\
\hline
 x   & y & y\\ 
y   & z & x\\  
z   & x & z     
\end{tabular}$$

But then, note that voters in profile $u^{III}_{(n)}$ with preferences $\begin{pmatrix} x \\ y \\ z \end{pmatrix}$ will misreport as $\begin{pmatrix} y \\ z \\ x \end{pmatrix}$ with a weight of $k$, and will misreport as $\begin{pmatrix} y \\ x \\ z \end{pmatrix}$ with a weight of $m$.  This will induce profile $u^{III}_2$, resulting in improved outcome $y$ over $z$.  This would contradict WSP, meaning that \textbf{Case 3.III.2.3.n+1} fails to hold in general.

So, we see that there is no voting rule $F$ other than the Condorcet method that satisfies the axioms P, A, N, and WSP on a rich domain.  This concludes \textbf{Step 3}, and concludes the proof of the theorem.

\newpage

\section*{Acknowledgements}

First and foremost, I thank my advisor Professor Eric Maskin for suggesting this incredible thesis topic, and for his guidance, patience, and all of his help throughout this project.  I am also grateful for the intellectually fascinating environment and courses of the Mathematics Department, and in particular, for the support of Professors Clifford Taubes and Dusty Grundmeier.  Finally, thank you to Professors Jerry Green, Christopher Avery, and Scott Kominers for their enlightening courses and for all of their advice and encouragement of my pursuit of economics.

\newpage

\section*{References}
\small{\noindent Arrow, Kenneth J.  \textit{Social Choice and Individual Values.} 1951.  John Wiley \& Sons, Inc.
\smallskip

\noindent Barbie, Martin, Clemens Puppe, and Attila Tasn\'adi.  2006.  ``Non-manipulable domains for the Borda count." \textit{Economic Theory.} 27(2):411:430.

\smallskip

\noindent Borda, Jean-Charles de. 1781. ``M\'emoire sur les \'elections au scrutin."  \textit{M\'emoire de l'Academie Royale des Sciences.} 657-665.

\smallskip

\noindent Condorcet, Marie Jean Antoine Nicolas Caritat.  1785.  \textit{Essai sur l'application de l'analyse \`a la pluralit\'e des voix}.  Imprimerie Royale.

\smallskip

\noindent Dasgupta, Partha and Eric Maskin. 2019. ``Elections and Strategic Voting: Condorcet and Borda." Working Paper.}

\end{document}